\documentclass[aps,prd,twocolumn,a4paper,groupedaddress,superscriptaddress,floatfix,nofootinbib]{revtex4-2}
\usepackage{graphicx}
\usepackage[urlcolor=blue,colorlinks,breaklinks=true]{hyperref}
\PassOptionsToPackage{obeyspaces,spaces,hyphens}{url}
\usepackage{amssymb}
\usepackage{amsmath}
\usepackage{hyperref}
\usepackage[english]{babel}

\begin{document}
\newcommand{\be}{\begin{equation}}
\newcommand{\ee}{\end{equation}}
\newcommand{\bq}{\begin{eqnarray}}
\newcommand{\eq}{\end{eqnarray}}
\newcommand{\bsq}{\begin{subequations}}
\newcommand{\esq}{\end{subequations}}
\newcommand{\bc}{\begin{center}}
\newcommand{\ec}{\end{center}}
\newcommand\lapp{\mathrel{\rlap{\lower4pt\hbox{\hskip1pt$\sim$}} \raise1pt\hbox{$<$}}}
\newcommand\gapp{\mathrel{\rlap{\lower4pt\hbox{\hskip1pt$\sim$}} \raise1pt\hbox{$>$}}}
\newcommand{\dpar}[2]{\frac{\partial #1}{\partial #2}}
\newcommand{\sdp}[2]{\frac{\partial ^2 #1}{\partial #2 ^2}}
\newcommand{\dtot}[2]{\frac{d #1}{d #2}}
\newcommand{\sdt}[2]{\frac{d ^2 #1}{d #2 ^2}}
\newcommand{\vv}[0]{{\bar v}}
\newcommand{\cc}[0]{{\tilde c}}
\newcommand{\ave}[1]{\left< #1 \right>}
\newcommand{\Ogw}[0]{\Omega_{\rm gw}}
\newcommand{\lp}[0]{{\rm f}}

\newcommand{\mA}[0]{\mathcal{A}}
\newcommand{\HH}[0]{\mathcal{H}}

\title{Gravitational waves from cosmic strings with friction: analytical approximations and parameter space}

\author{S. Mukovnikov}
\email[Electronic address: ]{Sergei.Mukovnikov@astro.up.pt}
\affiliation{Instituto de Astrof\'{\i}sica e Ci\^encias do Espa{\c c}o, Universidade do Porto, CAUP, Rua das Estrelas, PT4150-762 Porto, Portugal}
\affiliation{Centro de Astrof\'{\i}sica da Universidade do Porto, Rua das Estrelas, PT4150-762 Porto, Portugal}
\affiliation{Departamento de F\'{\i}sica e Astronomia, Faculdade de Ci\^encias, Universidade do Porto, Rua do Campo Alegre 687, PT4169-007 Porto, Portugal}

\author{L. Sousa}
\email[Electronic address: ]{Lara.Sousa@astro.up.pt}
\affiliation{Instituto de Astrof\'{\i}sica e Ci\^encias do Espa{\c c}o, Universidade do Porto, CAUP, Rua das Estrelas, PT4150-762 Porto, Portugal}
\affiliation{Centro de Astrof\'{\i}sica da Universidade do Porto, Rua das Estrelas, PT4150-762 Porto, Portugal}
\affiliation{Departamento de F\'{\i}sica e Astronomia, Faculdade de Ci\^encias, Universidade do Porto, Rua do Campo Alegre 687, PT4169-007 Porto, Portugal}

\begin{abstract}

We derive analytical approximations to describe the ultra-high-frequency secondary peak of the stochastic gravitational wave background generated by cosmic strings that is sourced by loops created in the friction-dominated era. We show that these approximations provide a very good description of the contribution of the friction-era loops over the relevant frequency range and for a broad range of cosmic string parameters, thus enabling a fast and accurate characterization of this signature. We also use these approximations to uncover the full parameter range in which this ultra-high-frequency peak should be distinguishable on the stochastic gravitational wave background spectrum and show that it should be present in a broader range of high-energy physics scenarios than originally reported in~\cite{Mukovnikov:2024zed}.
\end{abstract} 
\pacs{98.80.Cq}
\maketitle

\section{Introduction}

Ultra-high frequency Gravitational Waves (GWs) open the prospect of probing the very early universe directly and of helping uncover physics at very high energy scales (see ~\cite{Aggarwal:2020olq,Aggarwal:2025noe} for a review). Cosmic strings --- $1+1$-dimensional topological defects that may form in a plethora of early-universe scenarios~\cite{M_B_Hindmarsh_1995,Vilenkin:2000jqa,witten,jeannerot,sarangi} --- are a particularly promising source of GWs in this frequency range. Cosmic strings are expected to collide and interact frequently in their evolution and these interactions lead to the creation of closed loops of string that decay by emitting GWs. These emissions should then generate a Stochastic Gravitational Wave Background (SGWB) that spans a wide range of frequencies~\cite{Vilenkin:1981bx,Hogan:1984is,Accetta:1988bg} and whose amplitude and shape is sensitive not only to the underlying high-energy scenario~\cite{Blanco-Pillado:2013qja,Sousa:2013aaa,Sousa:2016ggw,Sousa:2020sxs,Hindmarsh:2021mnl,Buchmuller:2023aus,Rybak:2024our}, but also to the evolution of the cosmological background~\cite{Cui:2018rwi,Gouttenoire:2019kij,Auclair:2019wcv,Sousa:2020sxs,Blanco-Pillado:2024aca}.

The first stages of the evolution of cosmic string networks should be dominated by frictional forces caused by frequent interactions with the particles of the background plasma~\cite{Vilenkin:1981bx,Garriga}. This friction effectively decelerates cosmic strings and, as a result, it is typically assumed, in computations of the SGWB, that the contribution of loops created during the friction era is negligible. However, in earlier work~\cite{Mukovnikov:2024zed}, we have demonstrated that this is not necessarily the case: the production of loops during the friction era is so profuse that, although the individual emissions of loops are somewhat suppressed, in many instances their combined contribution may give rise to a prominent peak in the ultra-high frequency range of the spectrum.

Compellingly, in~\cite{Li:2025ybm}, the authors demonstrated that electromagnetic resonance systems should have the capability to probe the high-frequency cut-off of the SGWB generated by frictionless cosmic strings, which is precisely the frequency range where this friction peak may emerge. This opens the possibility of directly detecting this peak in the future and motivates further studies of the signatures of friction on the SGWB generated by cosmic strings. In this paper, we derive analytical approximations to describe the contribution of the loops created during the friction era to the SGWB. These approximations not only enable the fast characterization of the friction peak --- something that is very useful to derive/forecast observational constraints --- but also allow us to uncover the full parameter space in which one should expect this signature to be visible in this spectrum (which, as we shall see, is much broader than originally envisaged).

This paper is organized as follows. We start by reviewing the evolution of cosmic string networks and string loops during the friction era (Sec.~\ref{sec:evolution}) and the computation of the SGWB generated by string networks (Sec.~\ref{sec:SGWB}). In Sec.~\ref{sec:approx}, we derive analytical approximations to the contribution of loops created during the friction era to the SGWB and compare their results to a large number of numerically computed spectra, over a wide range of parameters, to verify their accuracy. We then use these approximations to derive analytical expressions to characterize the range of loop-size parameter wherein the signature of friction is distinguishable on the spectrum in Sec.~\ref{sec:parameter}. We then conclude in Sec.~\ref{sec:conc}.

Throughout this paper, we will use natural units with $c=\hbar=1$, where $c$ is the speed of light in vacuum and $\hbar$ is the reduced Planck constant. Moreover, we will use cosmological parameters determined using Planck 2018 data~\cite{2020A&A...641A...6P}, with the values of the density parameters for radiation, matter and dark energy at the present time respectively given by $\Omega_{\rm r} = 9.1476\times 10^{-5}$, $\Omega_{\rm m} = 0.308$, $\Omega_\Lambda=1-\Omega_{\rm r} - \Omega_{\rm m}$ and the Hubble constant by $H_0=2.13 h \times 10^{-33} \, \rm eV$, with $h=0.678$. 

\section{Evolution of cosmic strings with friction}\label{sec:evolution}

In this section, we briefly review the evolution of cosmic string networks and loops, focusing particularly on the friction-dominated early stages. For more details, see~\cite{M_B_Hindmarsh_1995,Garriga,Martins:1996jp,Vilenkin:2000jqa,Mukovnikov:2024zed}.

\subsection{Evolution of cosmic string networks}

The energy scale of the symmetry-breaking phase transition, $T_c$, at which the cosmic string network is produced determines the cosmic string mass per unit length, $\mu$, with $\mu \sim T_c^2$~\cite{Kibble:1976sj}. The corresponding formation time is roughly
\be
    \label{t_c}
    t_c = \frac{1}{\chi(t_c)} \frac{t_{pl}}{G \mu}, \quad\mbox{with}\quad \chi(t) = 4 \pi \left( \frac{\pi g_{*}(t)}{45} \right)^{1/2}\,,
\ee
where $t$ represents physical time, $G$ is the universal gravitational constant, $t_{pl}=G^{1/2}$ is the Planck time and $g_{*}(t)$ is the effective number of massless degrees of freedom (see e.g.~\cite{Kolb:1990vq}).

Right after the creation of the network, strings inevitably interact with the particles of the surrounding plasma and therefore experience a frictional force. The resulting force per unit length reads~\cite{PhysRevD.43.1060}
\be 
\mathbf{F}=-\frac{\mu}{l_f}\frac{\mathbf{v}}{\sqrt{1-v^2}}\,,
\ee 
where $\mathbf{v}$ is the string velocity and $l_f$ is the characteristic lengthscale above which friction significantly affects the evolution of the string. For gauge strings, the dominant contribution to friction arises from Aharonov-Bohm scattering~\cite{Alford:1988sj}, leading to the friction length of the form:
\begin{equation}
	\label{lf}
	l_f = \frac{\mu}{\beta(t) T^3}\,,
\end{equation}
where $T$ is the background temperature and $\beta$ is a parameter related to the number of effectively massless particle species that interact with a string (see e.g.~\cite{Martins:1996jp}) that is then, in general, time dependent.~\footnote{For simplicity, for the remainder of our paper, we will omit the dependency on time in the quantities that depend on $g_{*}(t)$ and/or the number of entropic degrees of freedom $g_S(t)$ (i.e., $\beta \equiv \beta(t)$, $\mathcal{G} \equiv \mathcal{G}(t)$, $\chi \equiv \chi(t)$, ...).}

The evolution of a cosmic string network can be described, on cosmological scales, using the Velocity-dependent One-Scale (VOS) model~\cite{Martins:1996jp,Martins:2000cs}. In this framework, the network is characterized by two variables: the Root-Mean-Squared (RMS) velocity $\vv$ of the network and the characteristic lengthscale $L$, which determines the average energy density of the network $\rho=\mu/L^2$. For standard local strings\footnote{This is not the case, for instance, for current-carrying strings~\cite{Martins:2020jbq} and strings with small-scale structure~\cite{Miranda:2026exq}.}, that typically may be treated as infinitely-thin and featureless, $L$ also provides a measure of the average distance between long strings. In this model, the evolution equations for these quantities are~\cite{Martins:1996jp,Martins:2000cs}
\bq
	\frac{d\vv}{dt} & = &  (1 - \vv^2) \left[ \frac{k(\vv)}{L} - \vv \left( 2H + \frac{1}{l_f} \right) \right]\,,\label{eq:VOSv}\\
	2\frac{dL}{dt} & = &  2HL(1 + \vv^2) + \frac{L \vv^2}{l_f} + \cc\vv\,,\label{eq:VOSL}
\eq
where $H=(da/dt)/a$ is the Hubble parameter and $a$ is the cosmological scale factor. Moreover, $k(\vv)$ is a phenomenological parameter that characterizes, to some extent, the impact of small-scale structure on long strings (here, we will use the \textit{ansatz} proposed in Ref.~\cite{Martins:2000cs}). $\cc=0.23\pm 0.04$~\cite{Martins:2000cs} is a phenomenological parameter that quantifies the impact of collisions and interactions between long strings. Typically strings are expected to intercommute in every interaction and this process may lead to the formation of closed loops of string that detach from the network. As a result, the network loses energy through the production of cosmic string loops.

In its evolution, a cosmic string network can pass through three distinct scaling regimes. The first two occur while friction dominates the dynamics, driving long strings to non-relativistic velocities. In fact, in the friction era, in which $l_f \ll 1/H$, Eq.~(\ref{eq:VOSv}) yields,  $\vv \simeq k_c l_f/L$, where $k_c\equiv k(t_c)\simeq k(0)=2\sqrt{2}/\pi$. During the first regime, known as the Stretching regime, strings are frozen in the cosmological background and are merely stretched by expansion. Quantitatively, in this case we have
\begin{equation}
    \label{L_s}
    L = L_c \left( \frac{t}{t_c} \right)^{1/2} \left( \frac{\mathcal{G}}{\mathcal{G}_c} \right)^{1/4}
\end{equation}
and
\be
    \label{v_s}
    \vv = \frac{\chi^{3/2}}{\chi_c^{1/2}} \frac{(G\mu)^{1/2}}{\beta} k_c \frac{t}{L_c} \left( \frac{\mathcal{G}_c}{\mathcal{G}} \right)^{1/4}\,,
\ee
where we use the subscript `$c$' to label the value of the corresponding quantity at $t=t_c$ and we have introduced
\be
    \label{G-factor}
    \mathcal{G}(t) \equiv \frac{g_*(t) g_S^{4/3}(t_0)}{g_*(t_0) g_S^{4/3} (t)}\,,
\ee
with $g_S$ denoting the effective number of relativistic entropic degrees of freedom, to account for the possibility that there is a change in the number of degrees of freedom during the friction era (a possibility that was not considered in~\cite{Mukovnikov:2024zed}). Notice that, typically, we should have $l_f (t_c)< L_c<t_c$~\cite{Martins:1996jp}.

The next regime, the Kibble regime, emerges when the Hubble damping term in Eq.~(\ref{eq:VOSL}) becomes comparable to the friction term. Quantitatively, this implies that during this regime
\be
    \label{Kibble_condition}
    AHL = \frac{l_f}{2L} k_c(k_c+\cc)\,,
\ee
with $A \simeq 3/2$~\cite{Mukovnikov:2024zed}. We assume that the transition between these regimes occurs suddenly at a time $t_k$, and that up to this moment the network follows the Stretching regime. We then define $t_k$ as the time at which the condition in Eq.~(\ref{Kibble_condition}), evaluated under the assumption of Eq.~(\ref{L_s}), is first satisfied. This corresponds to
\be
    \label{tk}
    t_k = \frac{t_{pl}}{\chi_k } \left( \frac{\mA \beta_k}{\chi_c (G \mu)^2} \right)^{2/3} \left( \frac{L_c}{t_c} \right)^{4/3} \left( \frac{\mathcal{G}_k}{\mathcal{G}_c} \right)^{1/3}\,,
\ee
where we have defined
\be
\mathcal{A} \equiv \frac{A}{k_c(k_c+\cc)} \simeq 1.454\,,
\ee
and we use the subscript `$k$' to label the values of the given quantity at $t=t_k$. After $t_k$, during the Kibble regime, Eq.~(\ref{Kibble_condition}) yields
\be
    \label{LKibble}
    L = \chi^{3/4}  \left( \frac{G\mu}{\mA \beta} \right)^{1/2} \frac{t^{5/4}}{t_{pl}^{1/4}}\,,
\ee
\be
    \label{vKibble}
    \vv = k_c \chi^{3/4} \left( \frac{\mA G\mu}{\beta} \right)^{1/2} \left(\frac{t}{t_{pl}}\right)^{1/4}\,.
\ee
Note that, during the Stretching regime, long strings hardly move with respect to each other and, therefore, do not interact. In contrast, during the Kibble regime, $HL \sim \cc \vv$, implying that the energy loss of the network due to loop production is comparable to that caused by friction and Hubble damping. Thus, as demonstrated in~\cite{Mukovnikov:2024zed}, loop production in the Kibble regime is significant. Notice also that, as manifest in~Eq.~\eqref{tk}, the duration of the Stretching regime depends on the initial characteristic length of the network. In fact, if condition~\eqref{Kibble_condition} is satisfied at $t_c$, the network will not go through the Stretching regime and it will start its evolution in the Kibble regime.

However, as the universe cools, friction weakens and its impact on cosmic string dynamics will eventually be negligible. Again, we assume a sudden transition into the frictionless era at a time $t_f$, in which the Hubble damping term $2H$ in Eq.~(\ref{eq:VOSv}) becomes equal to the friction damping term $l_f^{-1}$. This yields
\be
    \label{tf}
    t_f = \left( \frac{\beta_f}{G \mu} \right)^2 \frac{t_{pl}}{\chi_f^3} \,,
\ee
where the subscript `$f$' indicates that the quantities are evaluated at $t=t_f$.
\begin{figure}[t]
	\begin{center}
		\centering
		\includegraphics[width=\linewidth]{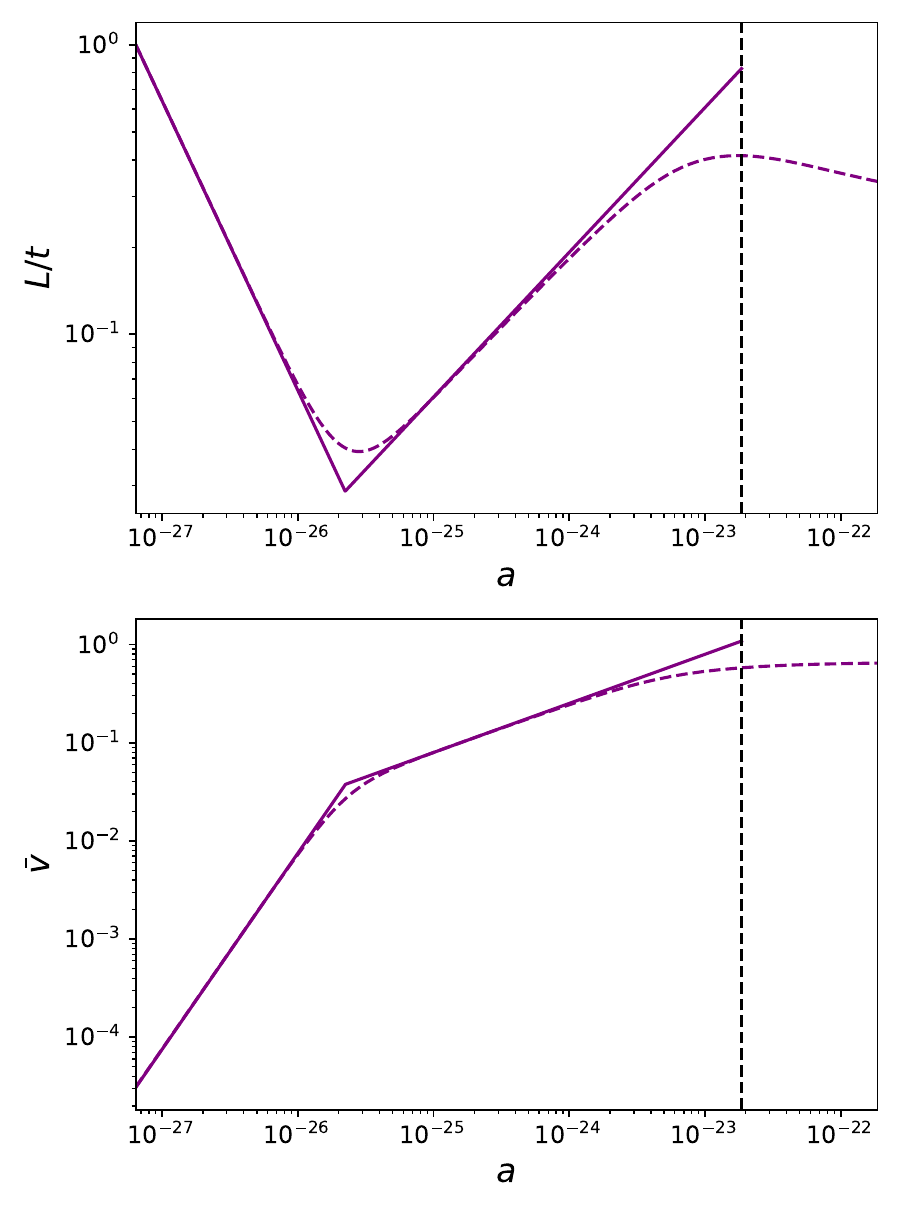}
		\caption{Evolution of a cosmic string network during the friction dominated era for $G\mu = 10^{-10}$, $\beta = 10$, $L_c = t_c$. The top panel displays the evolution of $L/t$, while the bottom panel displays $\vv$. The dashed lines correspond to the numerical calculation, while the solid lines represent the analytical approximations in Eqs.~(\ref{L_s}), (\ref{v_s}), (\ref{tk}), (\ref{LKibble}), (\ref{vKibble}) and (\ref{tf}). The vertical dashed line corresponds to the value of the scale factor at the end of the friction dominated epoch, $a_f$.}
		\label{fig:Full_evolution}
	\end{center}
\end{figure}

We display the evolution of $L/t$ and $\vv$ during these two friction-dominated regimes in Fig.~\ref{fig:Full_evolution}. The dashed and solid lines represent the numerical and analytical results, respectively. As may be seen, the analytical solution corresponding to Eqs.~(\ref{L_s}), (\ref{v_s}), (\ref{tk}), (\ref{LKibble}), (\ref{vKibble}) and (\ref{tf}) provides an excellent description of the network evolution, apart from the slow transitions between regimes.

After $t_f$, when string motion becomes frictionless, the network evolves towards a Linear scaling regime characterized by
\begin{equation}
	\label{2.6}
	L = \xi t, \;\; \frac{d\vv}{dt} = 0 \,,
\end{equation}
with
\be
    \label{linear}
    \xi^2 = \frac{k(k+\cc)}{4\nu(1-\nu)}, \;\; \vv^2 = \frac{k}{k+\cc} \frac{1-\nu}{\nu}\,,
\ee
where $\nu$ is defined by $a\propto t^{\nu}$. In this regime, string velocities are relativistic, resulting in frequent interactions and, similarly to the Kibble regime, significant loop production.

\subsection{Evolution of cosmic string loops}

The loops produced by the network are assumed to evolve independently from it and decay. In the frictionless case, their main decay mechanism is GW emission~\footnote{Note, however, that there may be other decay channels, depending on the particular cosmic string model. For instance, Abelian-Higgs cosmic strings may also emit scalar and gauge radiation, but the significance of these decay channels remains a matter of debate~\cite{Hindmarsh:2017qff,Hindmarsh:2021mnl,Blanco-Pillado:2022rad,Blanco-Pillado:2023sap}.} and this gives rise to an SGWB. However, during the friction era, interactions with the surrounding particles may, in principle, also affect the evolution of loops and their GW emission. As a matter of fact, we found in~\cite{Mukovnikov:2024zed} that the loops created during the Stretching always provide a negligible contribution to the SGWB. During the Kibble regime, however, although there is indeed a suppression of Gw emission, loops may still emit a significant amount of gravitational radiation. In reality, any loop will eventually reach a length below which their evolution may be regarded as frictionless and, from that point on, they decay only by emitting GWs, generating then a non-vanishing contribution to the SGWB.

During the Kibble regime, the evolution of the length of loops is well described by~\cite{Mukovnikov:2024zed}
\be
	\label{l}
	\ell = \ell_b exp\left[\vartheta \left( t^{-1/2} - t_b^{-1/2} \right) \right] - \Gamma G \mu \left( t - t_b \right)\,,
\ee
where $t_b$ and $l_b$ are the time and length at the time of birth of the loop, respectively, and $\vartheta = t_f^{1/2}$ if $g_*(t_b) = g_*(t) = g_*(t_f)$. Otherwise, assuming $g_*(t_b) = g_*(t)$, we have
\be
    \label{vartheta}
    \vartheta = \frac{\beta}{G\mu} \frac{t_{pl}^{1/2}}{\chi^{3/2}}\,.
\ee
The approximation in Eq.~(\ref{l}) takes into account both the effects of friction and of GW emission and it is highly accurate, in particular, when friction is strong. In the regime in which the effect of friction is comparable to that of GW emission, Eq.~(\ref{l}) leads to a slightly faster evaporation of loops, which translates into a slight underestimation of their contribution to the SGWB, yielding then conservative (but safe) results. We will then, for the sake of efficiency, use this approximation in our numerical computation of the SGWB. 

\section{The stochastic gravitational wave background generated by cosmic string networks} \label{sec:SGWB}

In this section, we review the computation of the SGWB generated by cosmic string networks and the contribution of friction-era loops to this spectrum.

\subsection{Computation of the spectrum} \label{sec:SGWB_calculation}

Loops are the main source of the SGWB produced by cosmic strings. Under the effect of their tension, they move and oscillate relativistically and, therefore, emit GWs. The superimposition of all their emissions constitutes the SGWB~\cite{Vilenkin:1981bx,Hogan:1984is,Accetta:1988bg}. We characterize 
this background as the energy density of gravitational radiation, ${\rho}_{\rm gw}$, per logarithmic frequency interval, normalized by the critical density of the universe at the present time $t_0$, $\rho_{\rm crit}$:
\begin{equation}
    {\Omega}_{\rm gw}(f)=\frac{1}{\rho_{\rm crit}}\frac{d{\rho}_{\rm gw}}{d\log f}\,,
\end{equation}
where $\rho_{\rm crit}=3H_0^2/(8\pi G)$, and, for the remainder of this paper, the subscript `$0$' indicates that a quantity is evaluated at the present time.

Loops emit GWs in harmonics of their length $\ell$. The frequency of the $j$-th harmonic mode, emitted at time $t$ and observed at $t_0$, is~\cite{Burden:1985md,Allen:1991bk}
\begin{equation}
	\label{f}
    f_j = \frac{2j}{\ell(t)} \frac{a(t)}{a_0}\,.
\end{equation}
Furthermore, we use the subscript `$j$' to label the contribution of the $j$-th mode of emission to the corresponding quantity. The distribution of power in the different harmonics approximately follows a power law of the form~\cite{Burden:1985md,Garfinkle:1988yi,Allen:1994bs,Damour:2000wa,Blanco-Pillado:2015ana}
\begin{equation}
    \frac{dE_j}{dt}=\Gamma_j G\mu^2\,, \quad \mbox{with}\quad \Gamma_j=\frac{\Gamma}{\zeta(q)}j^{-q}\,,
    \label{eq:powerspectrum}
\end{equation}
where $E=\mu\ell$ is the energy of the loop, $\zeta(q)$ is the Riemann Zeta function and $\Gamma\sim 50$~\cite{Vachaspati:1984gt,Burden:1985md,Scherrer:1990pj,Quashnock:1990wv,Allen:1994bs,Blanco-Pillado:2015ana} is the GW emission efficiency. The spectral index $q$ depends on the type of the small-scale structure present in the cosmic string loops. In particular, $q=4/3$ corresponds to the presence of cusps, i.e. points on the loop where the velocity instantaneously reaches $1$; $q=5/3$ corresponds to discontinuities in the string that propagate along it at the speed of light, known as kinks; and $q=2$ corresponds to collisions between kinks.

The amplitude of the SGWB produced by cosmic strings is given by (see e.g.~\cite{Vilenkin:2000jqa}):
\begin{equation}
    \label{SGWB}
   {\Omega}_{\rm gw}(f)=\sum_{j=1}^{+\infty} \Gamma_j {\Omega}_{\rm gw}^j(f)\,,
\end{equation}
where
\begin{align}
    \label{SGWBj}
    {\Omega}_{\rm gw}^j(f) =\frac{16 \pi}{3f}\left(\frac{G\mu}{H_0}\right)^2 \int_{t_{i}}^{t_e} j\, n\left(\ell_j(t), t\right) \left( \frac{a(t)}{a_0} \right)^5dt \nonumber \\
\end{align}
is the contribution of the $j$-th harmonic mode of emission. The initial and final times of integration, $t_i$ and $t_e$, can be chosen according to the relevant period of GW emission. For instance, for the contribution of the loops created during the frictionless era, $t_i = t_f$ and $t_e = t_0$. $\ell_j(t)$ is the length of the loops that radiate at the time t, in the $j$-th harmonic mode, GWs that have a frequency $f$ at the present time.

Note that, given the fundamental mode of emission, with $j = 1$, the $j$-th mode can be straightforwardly constructed using the relation
\begin{equation}
    \label{SGWBjf}
    {\Omega}_{\rm gw}^j(jf)={\Omega}_{\rm gw}^1(f)\,.
\end{equation}
Given this, for simplicity, we will consider only the fundamental mode in this paper. Therefore, for the amplitude of the SGWB, instead of Eqs.~(\ref{SGWB}) and~(\ref{SGWBj}), we will use
\begin{align}
    \label{SGWB1st}
    {\Omega}_{\rm gw}(f) =\frac{16 \pi}{3f}\left(\frac{G\mu}{H_0}\right)^2 \Gamma \int_{t_{i}}^{t_e} \, n\left(\ell(t), t\right) \left( \frac{a(t)}{a_0} \right)^5dt\,, \nonumber \\
\end{align}
which corresponds to assuming that all energy is radiated in the fundamental mode. Here the loop distribution function, $n(\ell,t)d\ell$, quantifies the number density of string loops with lengths between $\ell$ and $\ell+d\ell$ that exist at the time $t$:
\begin{equation}
    \label{numberdensity_int}
    n(\ell,t) = \int_{t_c}^{t} dt_b f(\ell_b, t_b) \left( \frac{a(t_b)}{a(t)} \right)^3\,,
\end{equation}
where $f(\ell, t)$ is the loop production function. Accordingly, $f(\ell, t)d\ell$ measures the number density of loops with lengths between $\ell$ and $\ell + d\ell$ produced per unit time. Moreover, to find the time of birth $t_b$ of a loop that has length $\ell$ at time $t$, we use Eq.~(\ref{l}).

The energy lost by the network due to loop production is of the form~\cite{Kibble:1984hp}:
\be 
    \label{eq:loss}
    \left. \frac{d\rho}{dt}\right|_{loops}=\cc \vv \frac{\rho}{L}\,,
\ee 
and, on the other hand, we should have that
\be 
    \label{eq:loss_int}
    \left. \frac{d\rho}{dt}\right|_{loops} = \mu \int_{0}^{\infty} \ell f(\ell, t)d\ell\,.
\ee 
Assuming (as is common in the literature) that the initial loop length is determined by the characteristic length $L$ at that time, with a constant loop-size parameter $\alpha$, such that
\begin{equation}
    \label{lb}
\ell(t_b) = \alpha L(t_b)\,,
\end{equation}
with $\alpha<1$, and using Eqs.~(\ref{eq:loss}) and~(\ref{eq:loss_int}), we obtain~\cite{Sousa:2013aaa,Auclair:2019wcv,Sousa:2020sxs}:
\begin{equation}
	\label{f_prod}
	f(\ell,t) = \frac{\cc}{\sqrt{2} \ell} \frac{\vv(t)}{L(t)^3} \delta(\ell - \alpha L)\,,
\end{equation}
where the $1/\sqrt{2}$ correction factor accounts for the redshifting of the peculiar velocities of loops~\cite{Vilenkin:2000jqa}~\footnote{This may be used to construct the loop production function for any distribution of loop lengths at birth; see e.g~\cite{Sanidas:2012ee,Sousa:2020sxs}.}. 

From Eqs.~(\ref{numberdensity_int}) and~(\ref{f_prod}), we find that the loop distribution function is given by~\cite{Sousa:2013aaa}:
\bq
    \label{number_dencity}
    n(\ell,t) = \frac{dt_b}{d\ell_b} \frac{\cc \vv\left( t_b \right)}{\sqrt{2} \alpha L\left( t_b \right)^4} 
    \left( \frac{a \left( t_b \right)}{a \left( t \right)} \right)^3 \,,
\eq
where $dt_b/d\ell_b$ may be found using the approximation in Eq.~(\ref{l}), 
\be
    \label{dt_b/dl_b}
    \frac{dt_b}{d\ell_b} = \left( \alpha \frac{dL}{dt} \bigg|_{t_b} + \frac{\alpha L(t_b)}{2 t_b^{3/2}}t_f^{1/2}  + \Gamma G \mu \right)^{-1}\,,
\ee
and the number of relativistic degrees of freedom is fixed to its value at $t_f$, so that $\vartheta = t_f^{1/2}$ in Eq.~(\ref{l}).

The friction era occurs early in the evolution of the network, as well as in cosmic history. This implies that, depending on the values of the parameters of the model, the lengthscales of the problem may be very small during these regimes. To avoid unphysical scales, we restrict ourselves to the emissions of loops created with lengths larger than the Planck length, $l_{pl}$, and assume that GW emission begins only once the gravitational backreaction scale, $\Gamma G\mu L$, is well defined. Quantitatively, this corresponds to the requirement that:
\be
    \label{cut-offs}
    \alpha L > l_{pl} \quad \mbox{and} \quad \Gamma G\mu L > l_{pl}\,.
\ee

\subsection{The signature of friction} \label{sec:signature}

\begin{figure}
		\centering
		\includegraphics[width=\linewidth]{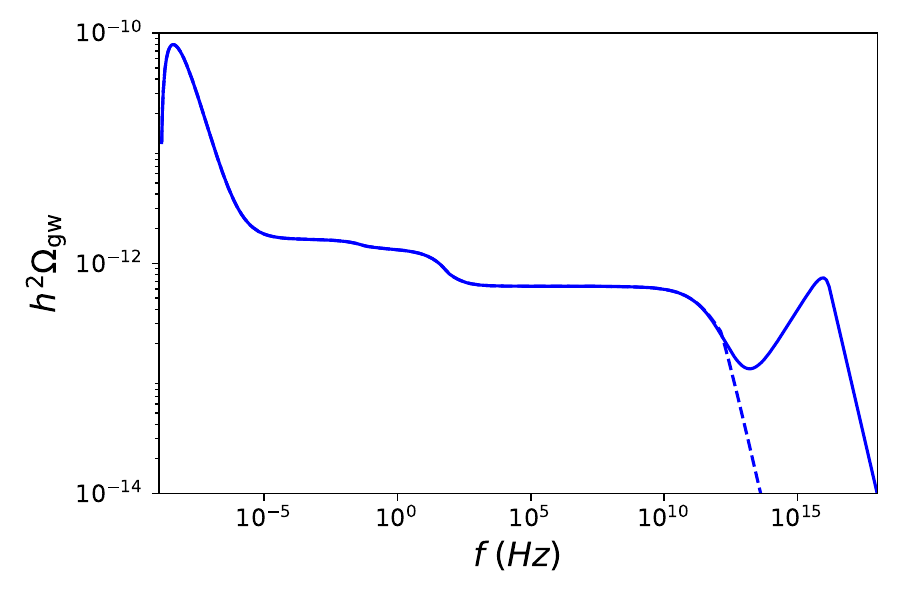}
		\caption{The Stochastic Gravitational Wave Background generated throughout the full evolution of a cosmic string network with $G \mu = 10^{-10}$, $\alpha = \Gamma G\mu$, $\beta = 10$, $L_c=t_c$. Here, the solid line represents the spectrum including the contribution of the loops created during the friction-dominated era of cosmic string network evolution, while the dashed line corresponds to the spectrum one obtains by assuming that significant GW emission by cosmic string loops starts only at the beginning of the frictionless regime.}
		\label{fig:SGWBfull}
\end{figure}

In earlier work~\cite{Mukovnikov:2024zed}, we used the method introduced in~\cite{Sousa:2013aaa} and briefly described in Sec.~\ref{sec:SGWB_calculation}, to fully characterize numerically the SGWB generated throughout the evolution of a cosmic string network, including the contribution of loops formed during the friction era (i.e., with $t_i = t_c$ and $t_e = t_0$).
 In Fig.~\ref{fig:SGWBfull}, we display an example of the resulting SGWB spectrum, including the often neglected contribution from the loops created in the friction era. Therein, it may be seen that the friction contribution extends the spectrum significantly toward the ultra-high frequency range and that the spectrum exhibits a secondary peak in this range. We found in~\cite{Mukovnikov:2024zed} that typically the main contribution to this signature is generated at the beginning of the Kibble regime. As a matter of fact, in the early stages of the Kibble regime, the network reaches its maximal density (in comoving coordinates), and, therefore, loop production is very intense, providing a prominent signature over significant ranges of parameter space, despite friction partially suppressing GW emission.

The signature of friction uncovered is sensitive to the size of loops, parametrized by $\alpha$, as well as to other cosmic strings parameters (namely $G\mu$, $L_c$ and $\beta$). Recall that these three parameters depend on the physics of the very early universe. Cosmic string tension $G\mu$ and the initial characteristic length $L_c$ are determined by the properties of the string-forming phase transition. The strength of friction, parametrized by $\beta$, depends not only on the underlying cosmic string model but also on the high-energy particle physics of the surrounding plasma, whose constituents interact (or not) with the strings. Note that, in the frictionless regime, the evolution of the string network does not depend on $L_c$ and $\beta$. Thus, the contribution generated during the friction era enables us to probe the physics of the early universe at energy scales that are inaccessible through the rest of the SGWB.

\section{Analytical approximations to the SGWB during the friction era}\label{sec:approx}

As was demonstrated in~\cite{Mukovnikov:2024zed}, we expect a prominent signature of friction in the SGWB not only in the small-loop regime, in which $\alpha \ll \Gamma G \mu$, but also for larger loops. In this section, we will derive analytical approximations to this spectrum in the small loop regime, as well as for loops created with a length of the order of the gravitational backreaction scale, $\Gamma G \mu L$ (that we will refer to as backreaction loops for short). We will later show that the approximation derived for backreaction loops remains valid for loops whose lengths are significantly larger than $\Gamma G \mu L$.

For the remainder of this paper, we will assume $g_*$ and $g_S$ remain constant during the Kibble regime. This is generally a good approximation if we consider the particle content predicted by the Standard Model of Particle Physics. In that case, provided that $G\mu > 3.7 \times 10^{-20}$, the friction era happens before the first decrease in the effective number of relativistic degrees of freedom. Thus, during the Kibble regime, we will fix the physical quantities dependent on $g_*$ and/or $g_S$ to their values at $t=t_k$ (i.e., $\beta = \beta_k$, $\chi = \chi_k$, $\mathcal{G} = \mathcal{G}_k$).

\subsection{Calculation} \label{sec:calculation}

Since the friction era takes place deep in radiation domination, to derive our analytical approximations, we will only retain the radiation component in the Friedmann equation. We will then approximate the Hubble parameter as\footnote{In the numerical computations of the spectra presented, we always consider a $\Lambda$CDM background and use Planck data to set the values of the cosmological parameters~\cite{planck}.}
\be
    \label{Friedmann}
    H(t)^2 = H_0^2 \Omega_{r} \mathcal{G}(t) \left( \frac{a_0}{a} \right)^4
\ee
and use the general formula for the SGWB in Eq.~(\ref{SGWB1st}). However, since the numerical computations in~\cite{Mukovnikov:2024zed} indicate that the signature of friction is generated mostly during the Kibble regime, we will neglect the contribution of the loops created during the Stretching regime and consider only the contribution of loops created at times $t_k\le t\le t_f$.

Since the characteristic length of the network grows during the Kibble regime, loops created in this epoch will be increasingly large and thus contribute to the SGWB at progressively lower frequencies (see Eqs.~\eqref{f} and~\eqref{lb}). In addition, according to Eq.~(\ref{f}), the lowest frequency emitted by a given loop is sourced at its time of creation, when its length is at its maximum.
So, at a given frequency $f$, the SGWB will only have contributions from loops created at times $t\ge t_{min}$, where $t_{min}$ is defined by~\cite{Sousa:2020sxs}
\begin{equation}
    \label{fa_min}
    f = \frac{2}{\ell_b(t_{min})} \frac{a(t_{min})}{a_0}\,.
\end{equation}
Using Eqs.~(\ref{LKibble}),~(\ref{lb}),~\eqref{Friedmann} and~(\ref{fa_min}), we find that this instant corresponds to a scale factor of
\be
    \label{amin}
    a_{min} = \left( \frac{4 \mA \HH^5}{t_{pl}^2} \right)^{1/3}\frac{\mathcal{G}_k^{5/12}}{(\alpha f)^{2/3} \chi_k^{1/2}}
    \left( \frac{\beta_k}{G \mu} \right)^{1/3}\,,
\ee
where $\HH \equiv \left( 2 H_0 \sqrt{\Omega_r} t_{pl} \right)^{1/2}$.
    
Therefore, to compute the SGWB spectrum, we start the integration in Eq.~(\ref{SGWB1st}) at a time $t_{min}$, when the first loops relevant for the given frequency $f$ were created. However, since we are only considering the contribution of the Kibble regime, whenever $t_{min} < t_k$ we take $t_k$ instead. Nonetheless, in such a case, this value needs to be further corrected due to the following. As may be seen from Fig.~\ref{fig:Full_evolution}, the transition between the Kibble and Stretching regimes happens slowly. This leads to an underestimation of the characteristic lengthscale at $t_k$ when using the approximation in Eq.~(\ref{LKibble}), relative to the numerical calculation. We found that starting the integration at a later time --- characterized by a scale factor of $\tau a_k$, with $\tau=1.44$ in the small-loop regime and $\tau=1.37$ for backreaction loops --- compensates for this and provides a good fit to the numerically computed SGWB spectra. Moreover, we have to take the cut-offs introduced in Eqs.~(\ref{cut-offs}) into account. These conditions mean that we do not consider loops created before $a_{cut}^1$ or $a_{cut}^2$, corresponding respectively to the cut-offs on the size of loops at birth or on the backreaction scale, where
\be
    \label{a_cut1}
    a_{cut}^1 = \HH \mathcal{G}_k^{1/4} \left( \frac{\mA \beta_k}{\alpha^2 G \mu \chi_k^{3/2}} \right)^{1/5} \,,
\ee
\be
    \label{a_cut2}
    a_{cut}^2 = \HH \mathcal{G}_k^{1/4} \left( \frac{\mA \beta_k}{\Gamma^2 (G \mu)^3 \chi_k^{3/2}} \right)^{1/5} \,.
\ee
By bringing all these conditions together, we may finally define the beginning of the integration in Eq.~(\ref{SGWB1st}), $t_x$. In terms of the scale factor, this corresponds to
\be
    \label{ax}
    a_x = \max(a_{min},\, \tau a_k,\, a_{cut}^1, \, a_{cut}^2)\,.
\ee
We then need to carry out the integration in Eq.~(\ref{SGWB1st}) from $t_i = t_x$ until $t_e = t_f$.

Backreaction and small loops evaporate quickly enough, compared to cosmological timescales, to assume, when computing the spectrum, that they evaporate effectively immediately after formation~\cite{Sousa:2014gka}. Therefore, we may assume that the time of birth of a loop, $t_b$, roughly coincides with the time of GW emission, $t$, and set $a(t)=a(t_b)$ in Eq.~\eqref{number_dencity}. The validity of this assumption will be demonstrated later in Sec.~\ref{sec:validation} through the comparison of the approximations developed here to the full numerical calculations. Furthermore, to perform the computation analytically, we will use, in Eq.~\eqref{number_dencity}, the analytical approximations to the evolution of the characteristic length $L$ and the RMS velocity $\vv$ during the Kibble regime in Eqs.~(\ref{LKibble}) and~(\ref{vKibble}).

\subsection{Approximations}\label{subsec:approx}

To perform the integral in Eq.~(\ref{SGWB1st}) analytically and derive approximations to the spectrum, we also need to simplify the Jacobian $dt_b/d\ell_b$ in Eq.~(\ref{number_dencity}). Let us start by considering small loops with $\alpha\ll\Gamma G\mu$. In this case, we can neglect the first two terms in the denominator of Eq.~(\ref{dt_b/dl_b}) and substitute only $\Gamma G \mu$ into Eq.~(\ref{number_dencity}). Note that the second term in Eq.~(\ref{dt_b/dl_b}) results from the effect of friction on loops. Thus, from the physical point of view, by neglecting it, we assume that in this regime the loops are not significantly affected by friction, which was demonstrated in~\cite{Mukovnikov:2024zed}. 

Under these simplifications, by performing the integral in Eq.~(\ref{SGWB1st}), we find:
\bq
    \label{SGWB_an_sm_loops}
    \Omega^{\rm sm}_{\rm gw}(f) = \frac{\mathcal{M}}{\alpha f} \frac{\beta_k^{3/2}}{(G \mu)^{1/2}} \frac{\mathcal{G}_k^{15/8}}{\chi_k^{9/4}}
    \left[ \frac{1}{a_x^{5/2}} - \frac{1}{a_f^{5/2}} \right]\,,
\eq
where
$$
    \mathcal{M} \equiv \frac{256 \pi k_c \cc \mA^{5/2} \HH^{7/2} \Omega_r}{15 \sqrt{2} t_{pl}} \,.
$$
From this approximation, it is straightforward to find the slopes of the SGWB. The left cut-off of the spectrum corresponds to the dominant emission from loops at the beginning of their evolution, which quantitatively means that $a_x = a_{min}$ in Eq.~(\ref{ax}). This generates a $f^{2/3}$ dependency.
The right end of the spectrum is generated by the subdominant emission from all loops in their end stages of the evolution, when they are effectively small, and, as in the standard spectrum it scales as $1/f$.

Let us now derive an analytical approximation to the spectrum generated by loops whose lengths are comparable to the gravitational  backreaction scale, with $\alpha \sim \Gamma G \mu$.
For $\alpha = \Gamma G\mu$ and at the time the dominant contribution to the background is generated, the terms in the denominator of Eq.~(\ref{dt_b/dl_b}) satisfy
\be
    \label{backr_estim}
    \alpha \frac{dL}{dt} \bigg|_{t_b} \lesssim  \Gamma G \mu \lesssim \frac{\alpha L(t_b)}{2 t_b^{3/2}}t_f^{1/2}
\ee
provided that $\beta$ is not too small (see App.~\ref{sec:estimations} for more details). This indicates that, in this regime, the evolution of loops cannot, unlike for small loops, be assumed to be frictionless. Given this, we may start by keeping just the second term in Eq.~(\ref{dt_b/dl_b}):
\be
    \label{dt_b/dl_b_simple}
    \frac{dt_b}{d\ell_b} \simeq \left( \frac{\alpha L(t_b)}{2 t_b^{3/2}}t_f^{1/2} \right)^{-1}\,.
\ee
In this case, after performing the integral in Eq.~(\ref{SGWB1st}), we obtain the following analytical approximation to the SGWB spectrum generated by backreaction loops born during the Kibble regime:
\bq
    \label{SGWB_an_back_simple}
    \Omega^{\rm br}_{\rm gw}(f) = \mathcal{N} \frac{G \mu \beta_k}{\alpha^2 f} \frac{\mathcal{G}_k^{5/2}}{\chi_k^3} \frac{g_S \left( a_k\right)}{g_S \left( a_0 \right)}
    \left( \frac{1}{a_x^2} - \frac{1}{a_f^2} \right)\,, 
\eq
where
$$
    \mathcal{N} \equiv \frac{64\sqrt{2} \pi k_c \cc \mA^3 \HH^6 \Gamma \Omega_r}{3 T_0^3 t_{pl}^4} \,
$$
and $a_x$ is given by Eq.~(\ref{ax}).
In this case, we find that the left and right cut-offs of the spectrum scale as $f^{1/3}$ and $1/f$ respectively.

\subsection{Validation of Approximations} \label{sec:validation}

We tested the analytical approximations in Eqs.~(\ref{SGWB_an_sm_loops}) and ~(\ref{SGWB_an_back_simple}) thoroughly by comparing their results to $450+$ numerically-computed spectra in the following range of the parameters\footnote{While considering cases with $G \mu = 10^{-20}$, we restricted the range of $\beta$ to [0.1,50] because of the emergence of numerical problems while solving the VOS equations for very small values of $G \mu$. For the same reason, for this value of tension, we only consider $L_c/t_c=1$ when $\beta=50$.
This however does not spoil the verification process of the approximations in Eqs.~(\ref{SGWB_an_back_simple}) and~(\ref{SGWB_an_back_more}), since, as we shall see, they typically work better for larger $\beta$ and/or smaller $L_c$.}:
\bq
    \label{ranges_of_parameters}
    \alpha \in [5 \cdot 10^{-24}, \, 0.1]\,, \; G \mu \in [10^{-7}, \, 10^{-20}]\,, \\ 
    \beta \in [0.1,\, 100]\,, \; L_c/t_c \in [0.01, \, 1]\,, \nonumber
\eq
and assuming the particle content predicted by the Standard Model of Particle Physics. Illustrative examples of such comparison are presented in Figs.~\ref{fig:SGWB_analytically_small_1} -~\ref{fig:SGWB_analytically_appendix}. In each of these figures, we fix the values of $G\mu$, $\beta$ and $L_c$, and vary the loop-size parameter $\alpha$. The contributions from the friction and frictionless eras, computed numerically, are represented by the solid and dash-dotted lines, respectively. The analytical approximations for small and backreaction loops are represented by the dotted and dashed lines respectively. In the following subsections, we discuss the quality of these approximations.

\subsubsection{Small loops}
\label{SGWB_small_validation}

\begin{figure} [t]
     \centering
     \includegraphics[width=0.9\linewidth]{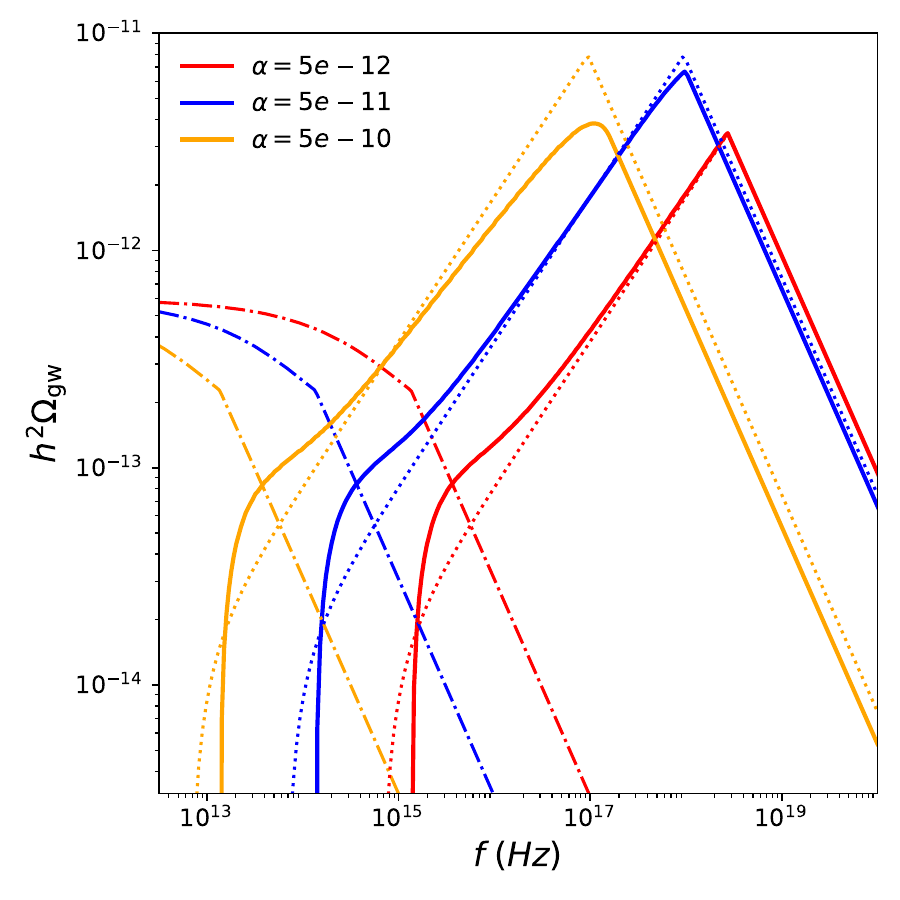}
     \caption{Analytical approximation to the SGWB generated by cosmic string loops created during the friction dominated epoch in the small loop regime. Here, we took $G \mu=10^{-10}$, $\beta=10$, $L_c=t_c$ and considered different values of $\alpha \ll \Gamma G\mu$. The solid lines represent spectra computed numerically, while the dotted lines correspond to the analytical approximation in Eq.~(\ref{SGWB_an_sm_loops}). The dash-dotted lines represent the contribution from the frictionless period of the networks evolution.}
     \label{fig:SGWB_analytically_small_1}
\end{figure}

\begin{figure} [t]
     \centering
     \includegraphics[width=0.9\linewidth]{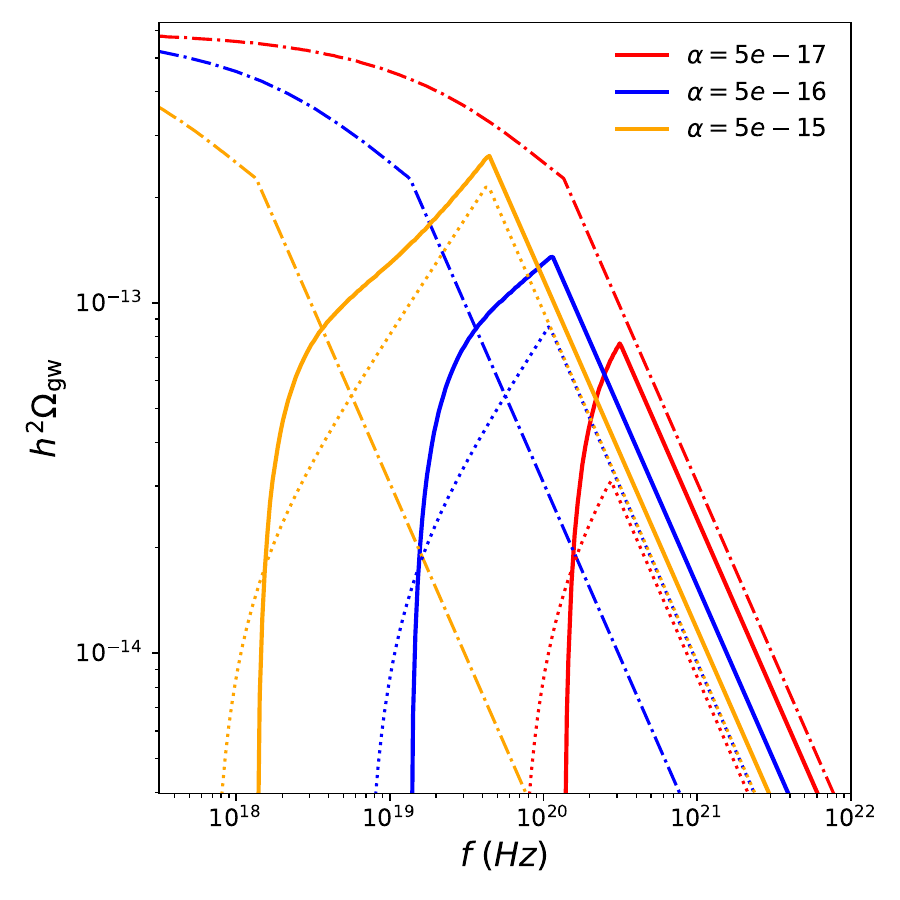}
     \caption{Analytical approximation to the SGWB generated by cosmic string loops created during the friction dominated epoch in the small loop regime. Here, we took $G \mu=10^{-10}$, $\beta=10$, $L_c=t_c$ and considered different values of $\alpha \ll \Gamma G\mu$. The solid lines represent spectra computed numerically, while the dotted lines correspond to the analytical approximation in Eq.~(\ref{SGWB_an_sm_loops}). The dash-dotted lines represent the contribution from the frictionless period of the networks evolution.}
     \label{fig:SGWB_analytically_small_2}
\end{figure}

\begin{figure} [t]
     \centering
     \includegraphics[width=0.9\linewidth]{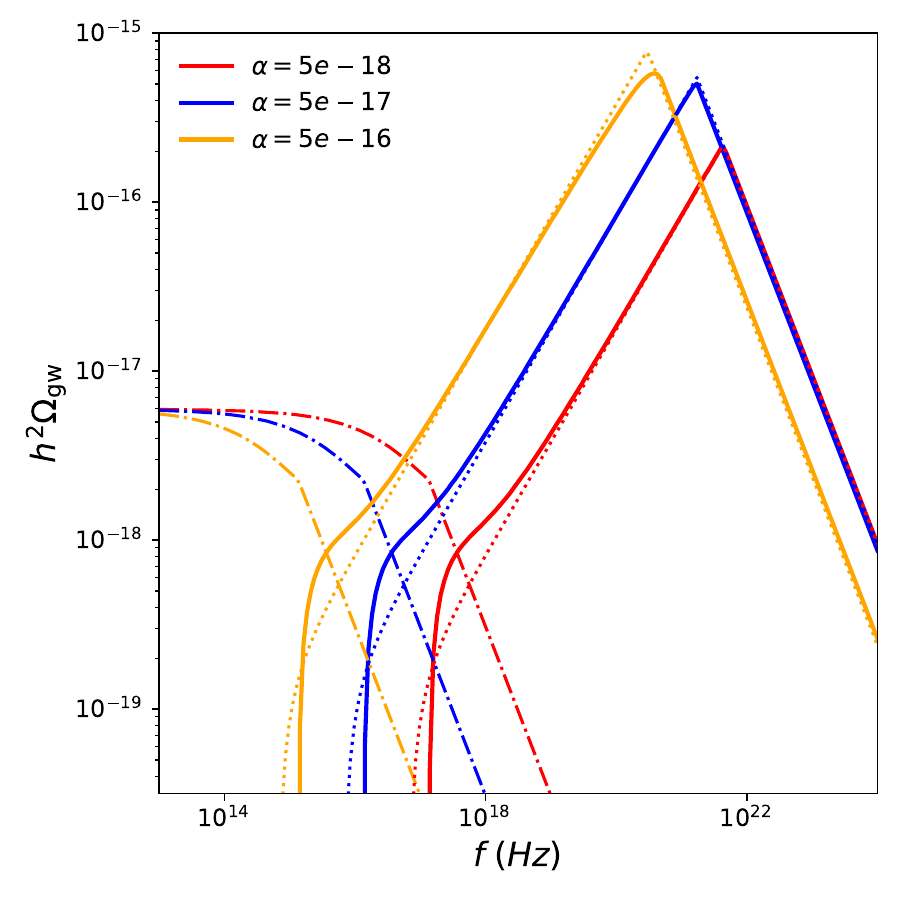}
     \caption{Analytical approximation to the SGWB generated by cosmic string loops created during the friction dominated epoch in the small loop regime. Here, we took $G \mu=10^{-15}$, $\beta=1$, $L_c=t_c$ and considered different values of $\alpha \ll \Gamma G\mu$. The solid lines represent spectra computed numerically, while the dotted lines correspond to the analytical approximation in Eq.~(\ref{SGWB_an_sm_loops}). The dash-dotted lines represent the contribution from the frictionless period of the networks evolution.}
     \label{fig:SGWB_analytically_small_3}
\end{figure}

When validating the approximation for small loops, for each set of parameters, the largest value of $\alpha$ considered was $\alpha=0.1\Gamma G\mu$ (represented by the orange line in Fig.~\ref{fig:SGWB_analytically_small_1}). We then gradually decreased loop size until the signature disappeared completely. This behavior is inevitable since, when we decrease $\alpha$, we necessarily encounter the cut-off $\alpha L > l_{pl}$ (i.e, $a_{cut}^1$ in Eq.~(\ref{a_cut1}) eventually reaches $a_f$). An example of $\alpha$ approaching this value is shown in Fig.~\ref{fig:SGWB_analytically_small_2}.

We verified that, in general, the approximation for the small loop regime provides excellent results over a very wide range of parameters, accurately predicting the shape, amplitude, and location of the signature in this regime. A typical example of such a fit is shown in Fig.~\ref{fig:SGWB_analytically_small_3}, where the nearly perfect agreement reflects the fact that the system lies deep within the small-loop regime. Moreover, one may see in Fig.~\ref{fig:SGWB_analytically_small_1} that this approximation also yields a good description of the signature in the transition regime, when $\alpha$ is one order of magnitude below the $\Gamma G\mu$ scale (the orange line on that figure).

This is, however, not always the case. The value of $\alpha$ below which the approximation for small loops works well is determined by the validity of the assumption that the motion of loops is frictionless at the beginning of integration. For the values of parameters considered, such $\alpha$ is roughly two orders of magnitude below $\Gamma G\mu$, which corresponds to the blue line in Fig.~\ref{fig:SGWB_analytically_small_1}. Furthermore, we found that this approximation works worse for larger $\beta$, corresponding to stronger friction 
and for smaller $G\mu$, as the motion of lighter strings is more affected by friction (this may be seen by comparing Figs.~\ref{fig:SGWB_analytically_small_1} and~\ref{fig:SGWB_analytically_small_3}). In these situations, the maximum value of $\alpha$ below which this approximation works may drop even further.

Fig.~\ref{fig:SGWB_analytically_small_2} also shows that, in the limit of very small loops --- when only the loops created in the end stages of the friction era satisfy Eq.~\eqref{cut-offs} and contribute to the SGWB (i.e, $a_{cut}^1$ is approaching $a_f$) ---, the approximation begins to underestimate the amplitude of the signature. This happens because, at the end of the Kibble regime, the network starts transitioning into a new scaling regime, and the analytical approximation in Eq.~(\ref{LKibble}) is no longer valid. This underestimation is most pronounced for values of $\alpha$ close to the minimum value for which there is a signature, a regime in which the friction peak is less distinguishable because it is mostly covered by the contribution generated in the frictionless era. Note however that this affects the results in a very limited range of $\alpha$ that is not very relevant physically (since the few loops that contribute to the spectrum before $a_f$ are extremely small, with length not much larger than $l_{pl}$).

Moreover, the low-frequency cut-off of the friction signature in our analytical approximations does not match the one calculated numerically. This cut-off is determined by the emission of the latest loop created during the friction era, at $t=t_f$. This discrepancy is caused by the fact that, as may be seen from Fig.~\ref{fig:Full_evolution}, the analytical approximation in Eq.~(\ref{LKibble}) overestimates the value of $L$ at $t_k$, leading to a lower minimum frequency. This, however, does not have a significant impact on the quality of our approximations since the lower end of the friction contribution is typically ``buried'' under the SGWB generated during the frictionless era.

\subsubsection{Backreaction and larger loops}
\label{SGWB_backreaction_validation}

\begin{figure} [t]
     \centering
     \includegraphics[width=0.9\linewidth]{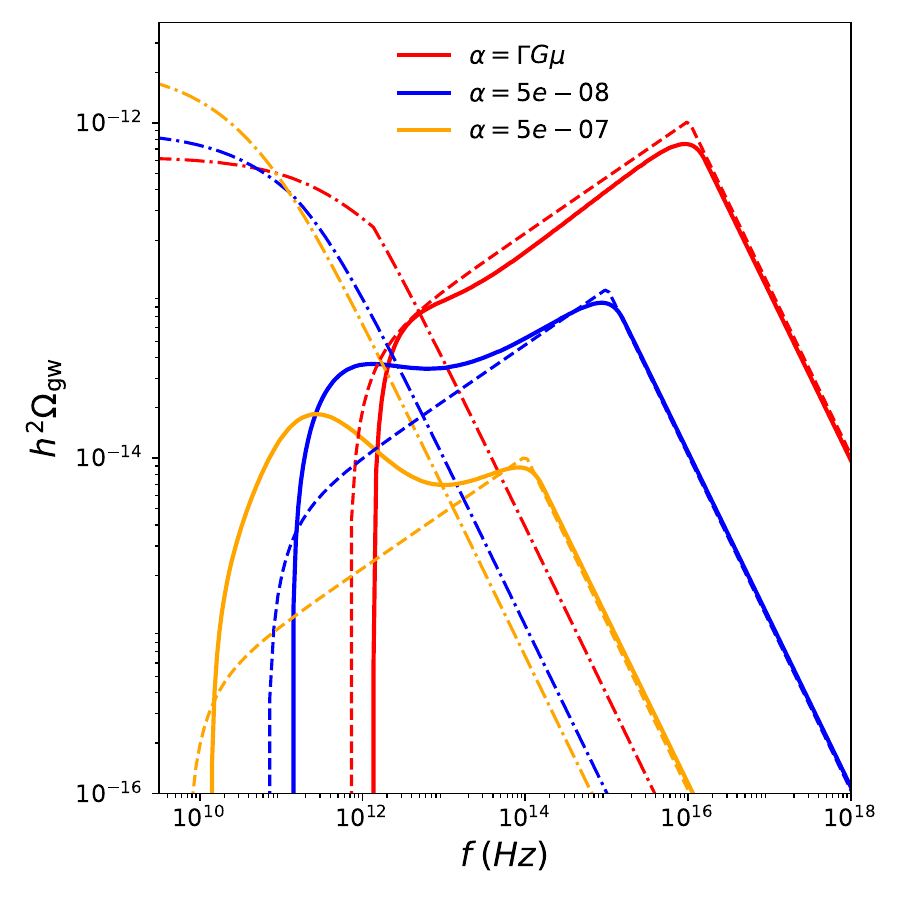}
     \caption{Analytical approximation to the SGWB generated by cosmic string loops created during the friction dominated epoch for backreaction and larger loops. Here, we took $G \mu=10^{-10}$, $\beta=10$, $L_c=t_c$ and considered different values of $\alpha\ge  \Gamma G\mu$. The solid lines represent spectra computed numerically, while the dashed lines correspond to the analytical approximation in Eq.~(\ref{SGWB_an_back_simple}). The dash-dotted lines represent the contribution from the frictionless period of the networks evolution.}
     \label{fig:SGWB_analytically_large_1}
\end{figure}

\begin{figure} [t]
     \centering
     \includegraphics[width=0.9\linewidth]{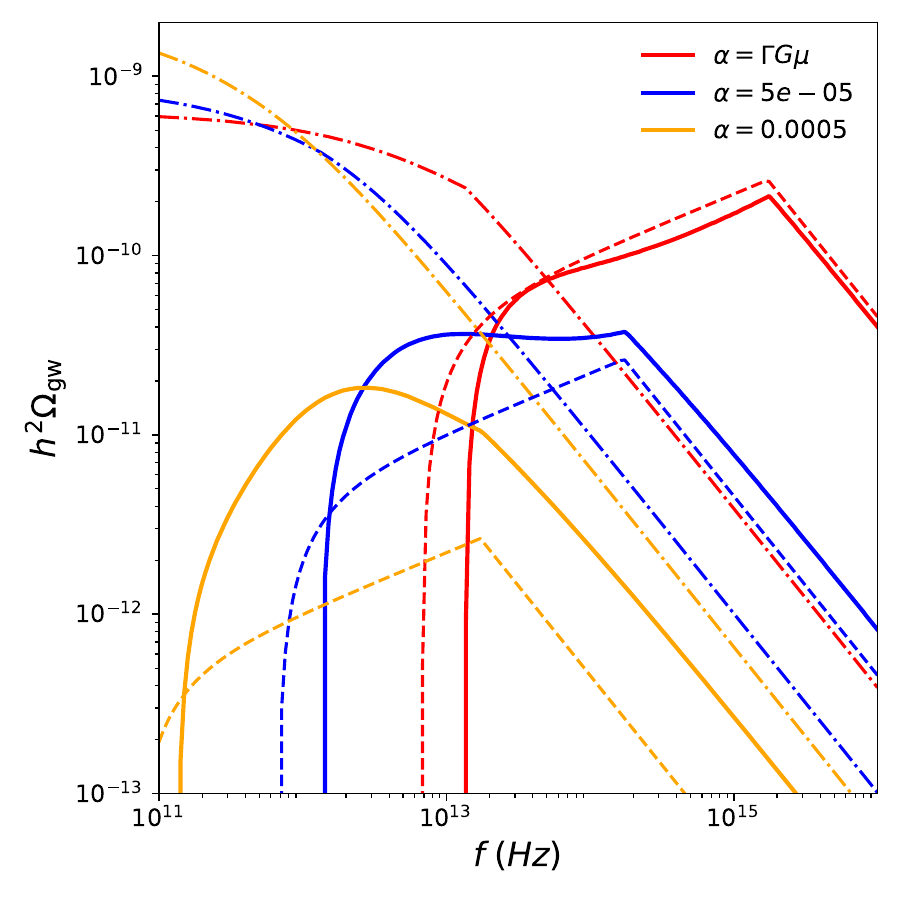}
     \caption{Analytical approximation to the SGWB generated by cosmic string loops created during the friction dominated epoch for backreaction and larger loops. Here, we took $G \mu=10^{-7}$, $\beta=1$, $L_c=0.01 t_c$ and considered different values of $\alpha\ge  \Gamma G\mu$. The solid lines represent spectra computed numerically, while the dashed lines correspond to the analytical approximation in Eq.~(\ref{SGWB_an_back_simple}). The dash-dotted lines represent the contribution from the frictionless period of the networks evolution.}
     \label{fig:SGWB_analytically_large_2}
\end{figure}

\begin{figure} [t]
     \centering
     \includegraphics[width=0.9\linewidth]{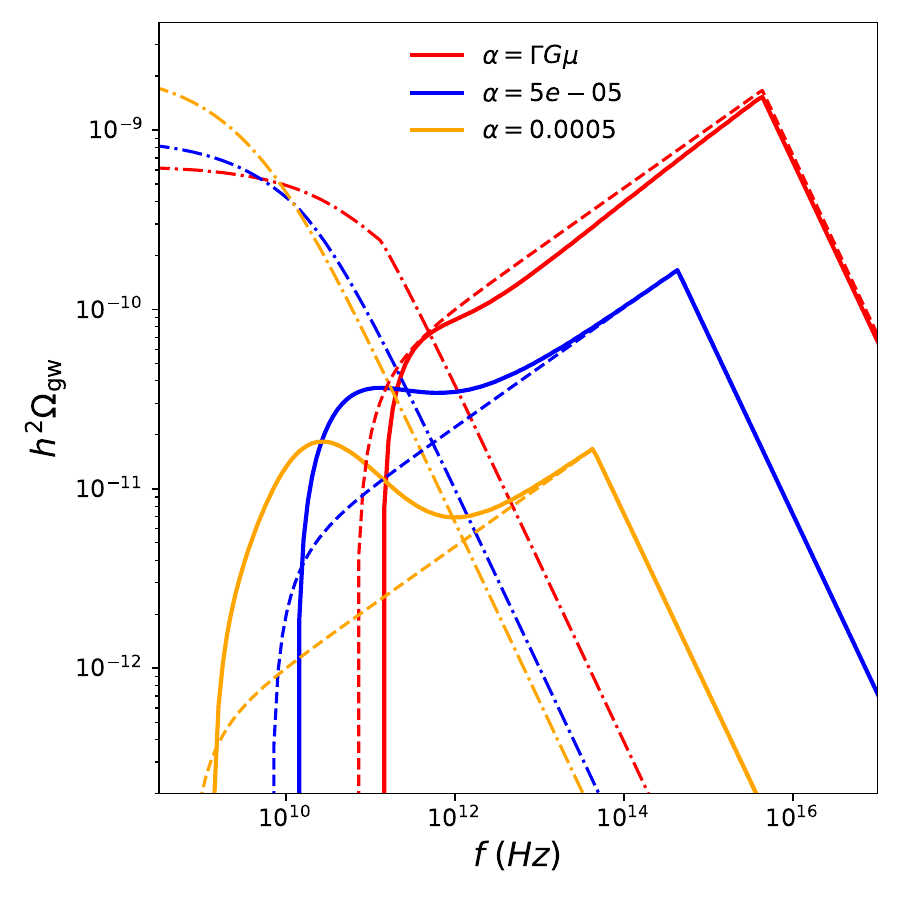}
     \caption{Analytical approximation to the SGWB generated by cosmic string loops created during the friction dominated epoch for backreaction and larger loops. Here, we took $G \mu=10^{-7}$, $\beta=100$, $L_c=0.01 t_c$ and considered different values of $\alpha\ge  \Gamma G\mu$. The solid lines represent spectra computed numerically, while the dashed lines correspond to the analytical approximation in Eq.~(\ref{SGWB_an_back_simple}). The dash-dotted lines represent the contribution from the frictionless period of the networks evolution.}
     \label{fig:SGWB_analytically_large_3}
\end{figure}

To validate the approximation for loops created with sizes comparable to the gravitational backreaction scale in Eq.~(\ref{SGWB_an_back_simple}) --- represented by the dashed lines in Figs.~\ref{fig:SGWB_analytically_large_1} -~\ref{fig:SGWB_analytically_large_3} ---  we started by considering loops with $\alpha=\Gamma G\mu$ (the red lines in Figs.~\ref{fig:SGWB_analytically_large_1} -~\ref{fig:SGWB_analytically_large_3}). We then moved to larger loops until the signature becomes smaller than the frictionless cut-off, which corresponds to the orange line in Fig.~\ref{fig:SGWB_analytically_large_2}. This behavior is inevitable as well, since larger loops are damped more effectively by friction and lose more energy as a result (rather than via GWs).

As demonstrated in Figs.~\ref{fig:SGWB_analytically_large_1} and~\ref{fig:SGWB_analytically_large_3}, this approximation proved to work very well, accurately predicting the slopes of the friction signature and the location of the peak not only for the case of backreaction loops (for which it was designed), but also for larger loops, up to values of $\alpha$ for which the signature becomes completely covered by the cut-off of the frictionless spectrum. We found, in our comprehensive analysis, that, although in some parameter ranges this approximation predicts a sharper peak than that obtained numerically (thus slightly overestimating the height of the peak), it typically performs better for larger $\beta$ and smaller $L_c$ (as may be seen in Fig.~\ref{fig:SGWB_analytically_large_3}). Accordingly, we expect this approximation to remain accurate outside the parameters ranges in Eq.~(\ref{ranges_of_parameters}) considered in this section. By contrast, smaller values of $\beta$ correspond to a weaker signature, rendering this region of parameter space less relevant.

In our analysis, we have also uncovered an additional peak in the signature of friction, in the low frequency range, that becomes more prominent for larger loops (see, for instance, in the orange lines in Figs.~\ref{fig:SGWB_analytically_large_1} and~\ref{fig:SGWB_analytically_large_3}). This bump appears to originate from the contributions of loops created late in the Kibble regime that do not satisfy the assumption of immediate evaporation. Although this additional contribution is not predicted by our approximation, this does not have a significant impact on the quality of our results since it is always mostly covered by the contribution generated after the end of the friction era. This lower frequency bump is at its most visible for $\alpha \sim \mathcal{O}(10\,\Gamma G\mu)$, which corresponds to the blue lines in Figs.~\ref{fig:SGWB_analytically_large_1} and~\ref{fig:SGWB_analytically_large_2}. In the latter (the worst such case uncovered), the amplitude of the peak of the spectrum is only underestimated by a factor of roughly $1.4$.

\subsubsection{Transitional regime}
Although, the two approximations derived here provide a good approximation to the SGWB generated by loops during the friction era for a wide range of parameters, we found a range of values of $\alpha$ in which they are not so accurate. For this reason, in App.~\ref{sec:more_general}, we derive a third analytical approximation. The starting point is the approximation for backreaction loops, but, in this case, we will include the second largest term ($\Gamma G\mu$) in Eq.~\eqref{dt_b/dl_b_simple} as well.
The analytical approximation we obtain in that case works well for the ranges of the parameters that are not covered by the approximations Eqs.~(\ref{SGWB_an_sm_loops}) and~(\ref{SGWB_an_back_simple}), but the expressions are more complex. We discuss this third approximation and its quality in App.~\ref{sec:more_general}.

\section{Parameter space of the friction signature}\label{sec:parameter}

The approximations derived in the previous section allow us to fully uncover the parameter space for which we have a signature of friction. The idea behind this is the following. If we consider a value of $\alpha$ that is too large, friction damps loops so efficiently that we do not get a distinguishable signature of friction in the SGWB spectrum. Then, if one decreases the length of loops, more energy is released in the form of GWs and the signature becomes prominent. This transition typically happens at a value of $\alpha$, that we will denote by $\alpha_{max}$, in which the contribution of friction to the SGWB is well described by Eq.~(\ref{SGWB_an_back_simple}). If we continue decreasing $\alpha$, we inevitably encounter the cut-off on the sizes of loops in Eqs.~(\ref{cut-offs}), which will eventually erase the signature completely. This happens for a value of $\alpha$, denoted by $\alpha_{min}$, that is typically in the range in which the small loops approximation Eq.~(\ref{SGWB_an_sm_loops}) is valid.
So, for $\alpha$ satisfying the condition
\be
    \label{alpha_sign}
    \alpha_{min} (G \mu, \, \beta, \, L_c) < \alpha < \alpha_{max} (G \mu, \, \beta, \, L_c)
\ee
there will be a distinguishable signature of friction in the ultra-high frequency range of the SGWB.

We may find $\alpha_{max}$ and  $\alpha_{min}$ by requiring that the amplitude of the SGWB generated during the friction era, $\Omega_{\rm gw}^{\rm fr}$, is larger than that generated in the frictionless period of evolution, $\Omega_{\rm gw}^{\rm lsc}$, at the peak frequency, $f_p$ of the friction contribution. Or equivalently, by requiring that:
\be
    \label{ratio_general}
    \frac{\Omega_{\rm gw}^{\rm fr} (f_p)}{\Omega_{\rm gw}^{\rm lsc} (f_p)} > 1\,.
\ee

\begin{figure*}[t]
    \centering
    \begin{minipage}{0.49\textwidth}
      \centering
      \includegraphics[width=\linewidth]{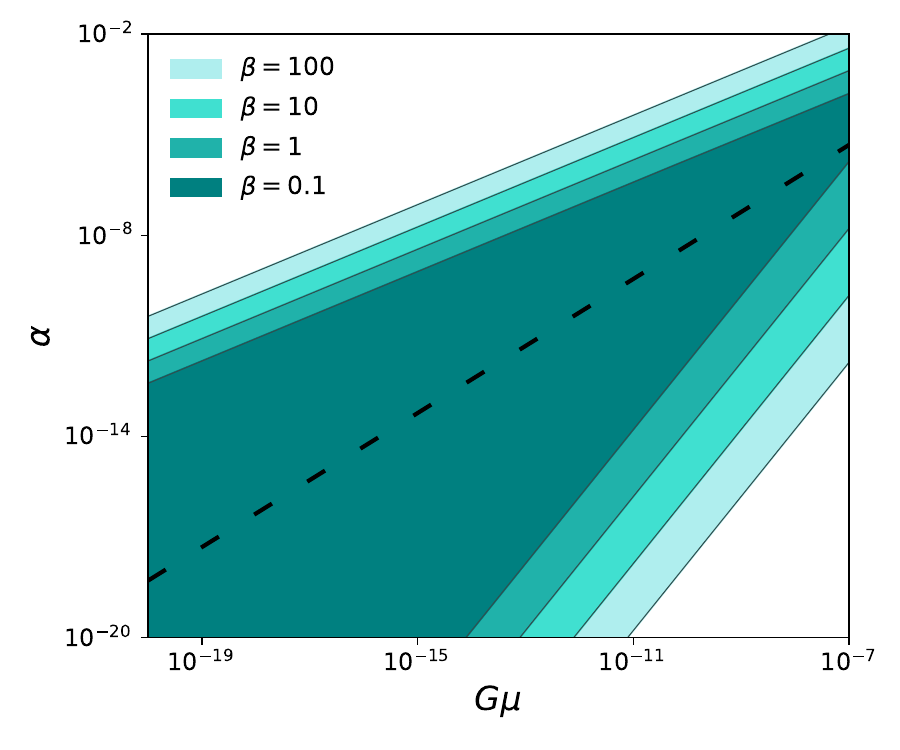}
    \end{minipage}\hfill
    \begin{minipage}{0.49\textwidth}
      \centering
      \includegraphics[width=\linewidth]{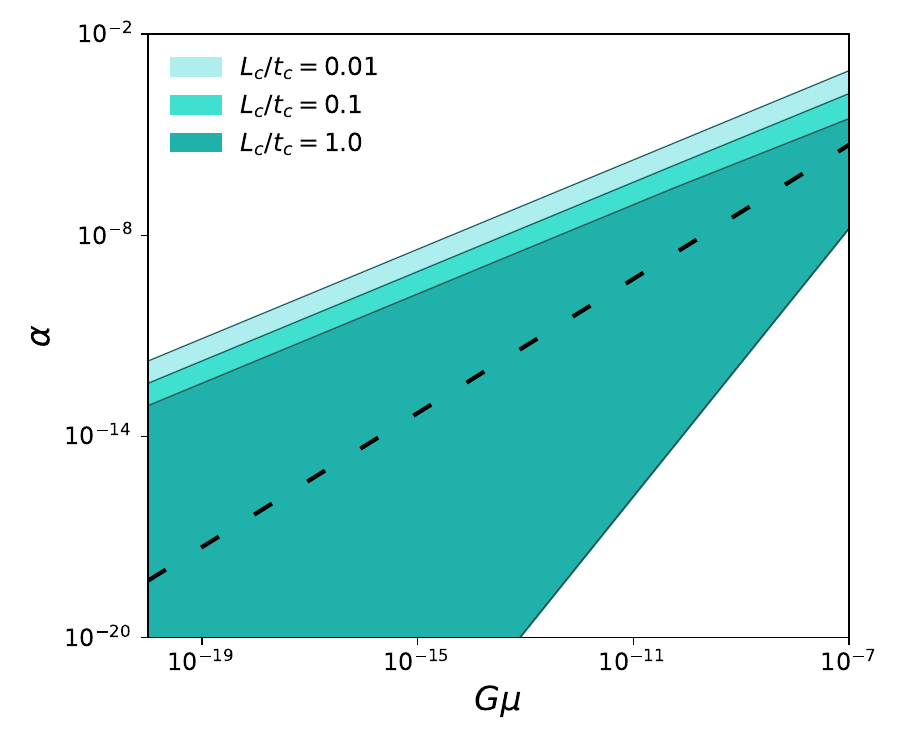}
    \end{minipage}
     \caption{Parameter space in which there is a distinguishable signature of friction in the ultra-high frequency range of the stochastic gravitational wave background. The left and right panels display the cases of fixed $L_c/t_c = 0.01$ and $\beta = 1$ respectively. The shaded areas with different colors represent the parameter ranges for which one will have a signature. The dashed line marks $\alpha = \Gamma G \mu$.}
     \label{fig:criterea}
\end{figure*}

The analytical approximations derived in the previous section provide an excellent fit for the location of the main frequency peak in the entire parameter space in Eq.~\eqref{ranges_of_parameters} (as may be seen in Figs.~\ref{fig:SGWB_analytically_small_1} -~\ref{fig:SGWB_analytically_appendix}).
From the discussion in the previous section, one may then infer that $f_p$ corresponds to a scale factor of
\be
    \label{fp_condition}
    a_p\equiv a_{min}(f_p) = \max(\tau a_k, \, a_{cut}^1, \, a_{cut}^2)\,.
\ee
The expression for the peak frequency then follows from Eq.~(\ref{f}) applied to the first mode of emission ($j = 1$) of the loops created at $a_p$:
\be
    \label{fp}
    f_p = \frac{2a_p}{\alpha L(a_p)} \,.
\ee

To find the amplitude of the frictionless cut-off in Eq.~(\ref{ratio_general}), we will use the analytical approximation derived in~\cite{Sousa:2020sxs}. Its simplified version reads
\begin{align}
    \label{Omega_gw_fr-less_cut-off}
    \Omega_{\rm gw}^{\rm lsc}(f) = \frac{128\pi}{3\sqrt{2\lambda}} \frac{\cc \vv_r}{\xi_r^4} \frac{\mathcal{H} \Omega_r}{t_{pl}}
    \frac{(G \mu)^2}{\alpha \beta_f f} \mathcal{G}_f^{5/4} \chi_f^{3/2} \left( 1 + \frac{\alpha \xi_r}{\Gamma G \mu} 
    \right)\,,
    \nonumber \\
\end{align}
where $\vv_r \simeq 0.66$, $\xi_r \simeq 0.27$ are, respectively, the scaling values of $\vv$ and $L/t$ during the radiation era, and we neglected the term resulting from the upper bound of integration since it provides a negligible contribution to the high-frequency cut-off of the SGWB generated during the frictionless era.
In this expression, we also included the effect of the decrease of the effective number of relativistic degrees of freedom and the parameter $\beta$~\footnote{As before, in this section we assume that the number of relativistic degrees of freedom remains constant during friction and immediately thereafter.}, which in~\cite{Sousa:2020sxs} was set to $1$. Moreover, while considering the full network evolution, we had to relax the assumption in~\cite{Sousa:2020sxs} that the network of cosmic strings is in the Linear scaling regime at the beginning of the frictionless stage of evolution. In a realistic setting, as illustrated in Fig.~\ref{fig:Full_evolution}, the transition between the two regimes is slow, and we do not expect the network to have already attained the Linear scaling regime by $t_f$. To correct for this effect, we have fitted the analytical approximation to the full numerical computation of the frictionless contribution, by starting the integration in Eq.~(\ref{SGWB1st}) at a later time $t_i=\lambda t_f$. We found that  $\lambda = 16$.

To estimate $\alpha_{max}$, we substitute the approximation in Eq.~(\ref{SGWB_an_back_simple}) and Eq.~\eqref{Omega_gw_fr-less_cut-off} into Eq.~(\ref{ratio_general}).
As in the case of the frictionless spectrum, we will neglect the second term in Eq.~(\ref{SGWB_an_back_simple}) since the main contribution to the signature comes from the earliest relevant period of time (at about $a_x$ from Eq.~(\ref{ax})). Also, we will assume that, for loops with $\alpha\ge\Gamma G\mu$, we do not encounter the cut-offs in Eqs.~(\ref{cut-offs}) and then take $a_p = \tau a_k$. Then, using Eq.~\eqref{fp}, we obtain obtain:

\begin{align}
    \label{alpha_max}
    & \alpha_{max} = \\ \nonumber
    & \frac{\Gamma G\mu}{2 \xi_r} \left( \sqrt{1 + p \left( \frac{\beta_k}{\sqrt{G\mu}} \frac{t_c}{L_c} \right)^{4/3} \frac{\mathcal{G}_k^{3/4}}{\chi_k^{17/6}} \frac{g_S(a_k)}{g_S(a_0)}} - 1 \right)\,,
\end{align}
where
$$
    p = 4 k_c \mathcal{A}^{7/3} \frac{\sqrt{\lambda}}{\tau^2}  \frac{\xi_r^5}{\vv_r} \left( \frac{\mathcal{H}}{T_0 t_{pl}} \right)^3\,.
$$

To estimate $\alpha_{min}$, we neglect the second term in Eq.~(\ref{SGWB_an_sm_loops}) (for the same reason as before). In addition, for $\alpha < \Gamma G \mu$, roughly speaking, we can neglect $\alpha \xi_r / \Gamma G \mu$ term in the brackets of Eq.~(\ref{Omega_gw_fr-less_cut-off}). To find the value of $\alpha$ below which the cut-offs Eqs.~(\ref{cut-offs}) erase the signature, we need to substitute $a_p = a_{cut}^1$ in Eq.~(\ref{fp}).
We then find
\be
    \label{alpha_min}
    \alpha_{min} = \frac{5}{2} \frac{\chi_k^3}{k_c \sqrt{\lambda}} \frac{\vv_r}{\xi_r^4} \left( \frac{G\mu}{\mathcal{A} \beta_k} \right)^2 \,.
\ee

To test the validity of these expressions for $\alpha_{min}$ and $\alpha_{max}$, we resorted to the $450+$ numerically-computed spectra used to validate the analytical approximations in Sec.~\ref{sec:validation}. We found that the condition in Eq.~\eqref{alpha_sign}, with $\alpha_{max}$ and $\alpha_{min}$ given by Eqs.~\eqref{alpha_max} and~\eqref{alpha_min}, accurately predicts the existence (or absence) of a signature in about $98.5\%$ of the cases, thus providing an excellent description of the parameter space in which there is a signature of friction. The cases in which this condition is not verified are caused by small discrepancies between the analytical value of $\alpha_{max}$ and its numerically determined value in the limits in which approximation in Eq.~\eqref{SGWB_an_back_simple} provides a less accurate fit to the main frequency peak. For instance, as illustrated in Fig.~\ref{fig:SGWB_analytically_large_1}, in some parameter ranges, this approximation predicts a sharper peak, which leads to a slight underestimation of $\alpha_{max}$ when using Eq.~\eqref{alpha_max}. In these situations (corresponding to about half of the $~1.5\%$ in which this test failed), we always have a signature of friction when the condition in Eq.~\eqref{alpha_sign} is verified, but this condition leaves out a (very small) part of the parameter space in which this signature is present. Moreover, as discussed in Sec.~\ref{sec:validation}, the approximation in Eq.~\eqref{SGWB_an_back_simple} does not predict the low-frequency bump in the friction signature that we have uncovered in the numerical computations and, as a result, it underestimates the height of the peak of the spectrum in the range of parameters in which this bump is at its most important (typically, when both $L_c/t_c$ and $\beta$ are small, but this depends on $G\mu$ as well). In this limit, we found that using Eq.~\eqref{alpha_max}, in some instances, leads to an overestimation of $\alpha_{max}$ and, as result, condition~\eqref{alpha_sign} predicts that there is a signature for a parameter range that is slightly broader than what we find numerically. Note however that this overestimation is not very significant: in the worst case we encountered, $\alpha_{max}$ was overestimated by a factor of about $3.5$.

Given this, we may use the analytical expressions for $\alpha_{max}$ and $\alpha_{min}$ to uncover the full parameter space in which a distinctive signature of friction should be expected on the SGWB generated by cosmic strings. This corresponds precisely to the regions of $\alpha$ that satisfy the condition in Eq.~\eqref{alpha_sign} and it is represented in Fig.~\ref{fig:criterea} by the shaded areas.
Note that the initial characteristic lengthscale, by causality, cannot exceed $1$. Moreover, one should have $\alpha < 1$ and that $G \mu \lesssim 10^{-7}$, as constrained by Cosmic Microwave Background data~\cite{Caloni:2026dyu,Raidal:2026cpb}. Fig.~\ref{fig:criterea} demonstrates the potential existence of a signature of friction for a wide range of cosmic string tensions, including scenarios in which the length of loops is significantly larger than the gravitational backreaction scale (the region above the dashed black lines). This is a significantly wider parameter space than that predicted in~\cite{Mukovnikov:2024zed}, which identified the signature for values of $\alpha$ that were not much larger than $\Gamma G\mu$. This figure also shows that this parameter space becomes significantly broader for larger $\beta$ and for smaller $L_c$.

\section{Conclusions}\label{sec:conc}

In this paper, we have derived analytical approximations to the ultra-high-frequency friction peak of the cosmic string SGWB spectrum that had previously been uncovered in~\cite{Mukovnikov:2024zed}. We have shown that these approximations provide a high-quality fit to this signature for a broad range of the $(G\mu,\alpha,\beta,L_c)$ parameter space, thus enabling a significantly faster characterization of this signature. At a time in which the ultra-high-frequency range of the GW spectrum is garnering increasing attention due to its potential as a probe of the physics of the very early universe and in which multiple GW detectors designs to probe this band are being proposed (see e.g.~\cite{Aggarwal:2020olq,Ito:2023fcr,Liu:2023mll,He:2023xoh,Vacalis:2023gdz,Aggarwal:2025noe}), these approximations may become a valuable tool. They should enable a fast evaluation of the ability of future detectors to probe cosmic string scenarios and for a quick forecast of the observational constraints that may be obtained through them. They may also be used in the future as templates to search for this signature observationally.

We have also used these analytical approximations to fully characterize the $(G\mu,\alpha,\beta,L_c)$ parameter range in which the secondary peak of friction should be distinguishable in the spectrum. This revealed that this signature should be present for a broad range of cosmic string scenarios, much broader than initially expected. Since the parameters $\beta$ and $L_c$ should be highly-dependent on the underlying particle physics framework~\cite{Mukovnikov:2024zed}, this indicates that the potential of the signature of friction as a probe of high-energy physics is larger than originally envisaged. In particular, although in~\cite{Mukovnikov:2024zed} this signature was shown to be present only for relatively small loops (i.e., loops with lengths that are not significantly larger than that the gravitational backreaction scale), here we show that it should also be present for larger loops as well. In fact, for $G\mu\sim \mathcal{O}(10^{-7})-\mathcal{O(}10^{-6})$, the secondary peak of friction may be present even for loop lengths as large as those predicted for Nambu-Goto cosmic strings ($\alpha\sim 0.34$). Note that, although current constraints resulting from pulsar-timing-array data limit the tension of cosmic string networks to $G\mu \lesssim10^{-11}$~\cite{EPTA:2023xxk,NANOGrav:2023hvm}, these stringent constraints may be evaded if strings are metastable and decay early enough in cosmic history~\cite{Buchmuller:2023aus} and, if they decay before the time of decoupling, they would also evade CMB constraints as well. The signature of friction may then provide us with a valuable alternative to probe observationally scenarios in which cosmic string networks are short-lived as well.

\acknowledgments

 S.M. is supported by FCT - Funda{\c c}\~ao para a Ci\^encia e a Tecnologia (\url{https://ror.org/00snfqn58}) through the PhD fellowship with reference UI/BD/152220/2021 (\url{https://doi.org/10.54499/UI/BD/152220/2021}). L.S. is supported by FCT through contract No 2024.07993.CEECIND/CIAAUP-14/2026/CTTI. This work was also funded by FCT through the research grant UID/04434/2025 (\url{https://doi.org/10.54499/UID/04434/2025}) and 2024.17828.PEX - \textit{Unveiling the early universe with topological defects} (\url{https://doi.org/10.54499/2024.17828.PEX}).

\appendix

\section{Evaluation of terms in the Jacobian} \label{sec:estimations}

To estimate the magnitude of the terms in the denominator of Eq.~(\ref{dt_b/dl_b}), let us take $\alpha = \Gamma G \mu$. In this case, we have
\be
    \label{dl_b/dt_b_backreaction}
    \frac{d\ell_b}{dt_b} =  \Gamma G \mu \left( \frac{dL}{dt} \bigg|_{t_b} + \frac{L(t_b)}{2 t_b^{3/2}}t_f^{1/2}  + 1 \right)\,.
\ee

Using the analytical approximation to the characteristic length of the network during the Kibble regime in Eq.~(\ref{LKibble}), we find
\bq
    \label{dLdtb}
    \frac{dL}{dt} \bigg|_{t_b} & = & \frac{5}{4} \chi^{3/4} \left( \frac{G\mu}{\beta\mathcal{A}} \right)^{1/2}
    \left( \frac{t_b}{t_{pl}} \right)^{1/4} \leqslant \\ \nonumber
    & \leqslant & \frac{5}{4} \chi^{3/4} \left( \frac{G\mu}{\beta \mathcal{A}} \right)^{1/2}  
    \left( \frac{t_f}{t_{pl}} \right)^{1/4} = \frac{5}{4\mathcal{A}^{1/2}}  \approx 1 \,,
\eq
where in the inequality we have considered the maximum possible value that this term can have. During the Kibble regime, this is achieved for $t_b = t_f$. Also, in Eq.~(\ref{dLdtb}), we used Eq.~(\ref{tf}). Thus, we have:
\be
    \label{dLdtb_constr}
    \frac{dL}{dt} \bigg|_{t_b} \lesssim 1\,,
\ee
where $1$ is achieved when we consider the emissions of loops created in the end of the Kibble regime, whose contribution to the spectrum, as was shown in~\cite{Mukovnikov:2024zed}, is subdominant.

Similarly, we may see that the second term in Eq.~(\ref{dl_b/dt_b_backreaction}), if one considers the times of births of the loops that provide the largest contribution to the signature of friction, $t_b=t_k$, takes the form
\be
    \label{2d_term_tk}
    \frac{L(t_b)}{2 t_b^{3/2}}t_f^{1/2} \bigg|_{t_b = t_k}  = \frac{1}{2\mathcal{A}^{2/3}}
    \frac{1}{(G \mu)^{1/6}} \left( \frac{\beta}{\chi} \frac{t_c}{L_c} \right)^{1/3}\,,
\ee
where we assumed $\chi(a_c) = \chi$ and $\mathcal{G}(a_c) = \mathcal{G}$.
Quite generally, we should have that
\be
   \label{2d_term_constr_2}
   \frac{L(t_b)}{2 t_b^{3/2}}t_f^{1/2} \bigg|_{t_b = t_k} \gtrsim 1\,,
\ee
provided that
\be
    \label{2d_term_condition}
    \frac{1}{(G \mu)^{1/6}} \left( \frac{\beta}{\chi} \frac{t_c}{L_c} \right)^{1/3} \gtrsim \frac{2}{\mathcal{A}^{2/3}} \simeq 2.57\,.
\ee
We expect this condition to hold then if $\beta$ is not too small. In the worst case scenario --- when the second term in~\eqref{dl_b/dt_b_backreaction} is at its minimum, for $L_c = t_c$ --- and assuming the Standard Model of particle physics, we find that condition~\eqref{2d_term_constr_2} should hold provided that: 
\be
    \label{beta_min}
    \beta \gtrsim 582 \sqrt{G\mu}\,.
\ee
Since we do not anticipate a prominent signature when friction is very weak (i.e. when $\beta \ll 1$), this condition should then hold for the relevant parameter ranges.

Combining Eqs.~(\ref{dLdtb_constr}) and~(\ref{2d_term_constr_2}), we find that, at $t_k$, when the dominant contribution to the signature of friction is generated, the terms of Eq.~(\ref{dt_b/dl_b}) satisfy:
\be
    \label{backr_estim_2}
    \alpha \frac{dL}{dt} \bigg|_{t_b} \lesssim  \Gamma G \mu \lesssim \frac{\alpha L(t_b)}{2 t_b^{3/2}}t_f^{1/2}\,,
\ee
for $\alpha = \Gamma G\mu$ and $\beta$ satisfying~\eqref{beta_min}.

\section{Analytical approximation for the transitional regime} \label{sec:more_general}

Although the approximations we have developed in Sec.~\ref{subsec:approx} work very well deep in the small-loop regime and for loops with lengths comparable to or larger than the gravitational backreaction scale, their results are less accurate for values of $\alpha$ in the transition between these two regimes. In this appendix, we derive a more complete and complex analytical approximation to describe this transitional regime. If we include the two largest terms in Eq.~(\ref{dt_b/dl_b}) --- which according to the estimate~(\ref{backr_estim}) are respectively $\alpha L(t_b)t_f^{1/2} / 2 t_b^{3/2}$ and  $\Gamma G \mu$ --- we may still perform the integral in Eq.~(\ref{SGWB1st}) analytically to obtain:
\begin{widetext}
\bq
    \label{SGWB_an_back_more}
    \Omega^{\rm tr}_{\rm gw}(f) = \mathcal{N}' \frac{1}{\alpha^6 f} \frac{(G \mu)^7}{\beta_k} \frac{\mathcal{G}_k^5}{\chi_k^6} 
    \left( \frac{g_S \left( a_k\right)}{g_S \left( a_0 \right)} \right)^5 \Bigg[ 2 \ln\left( \frac{\sqrt{a_x} +d}{\sqrt{a_f} +d} \right) +
    \ln \left( \frac{a_f}{a_x} \right) + \quad  \\ \nonumber
    + \frac{d^4}{2} \left( a_x^{-2} - a_f^{-2} \right) + \frac{2}{3} d^3 \left( a_f^{-3/2} - a_x^{-3/2} \right) +
    d^2 \left( a_x^{-1} - a_f^{-1} \right) + 2 d \left( a_f^{-1/2} - a_x^{-1/2} \right) \Bigg]\,,
\eq
where
$$
    \mathcal{N}' \equiv \frac{2^{11} \sqrt{2} \pi k_c \cc}{3} \left( \frac{\Gamma \mA}{T_0^3} \right)^5 
    \left( \frac{\HH}{t_{pl}} \right)^{16} \Omega_r \,, \;\;\;
    d \equiv \alpha \left(\frac{\beta_k}{(G\mu)^3 \mA \HH^5} \right)^{1/2} \frac{\chi_k^{3/4}}{\mathcal{G}_k^{5/8}}
    \frac{g_S \left( a_0 \right)}{g_S \left( a_k\right)} \frac{\left(T_0 t_{pl} \right)^3}{2 \Gamma} \,.
$$
\end{widetext}

\begin{figure}
     \centering
     \includegraphics[width=0.9\linewidth]{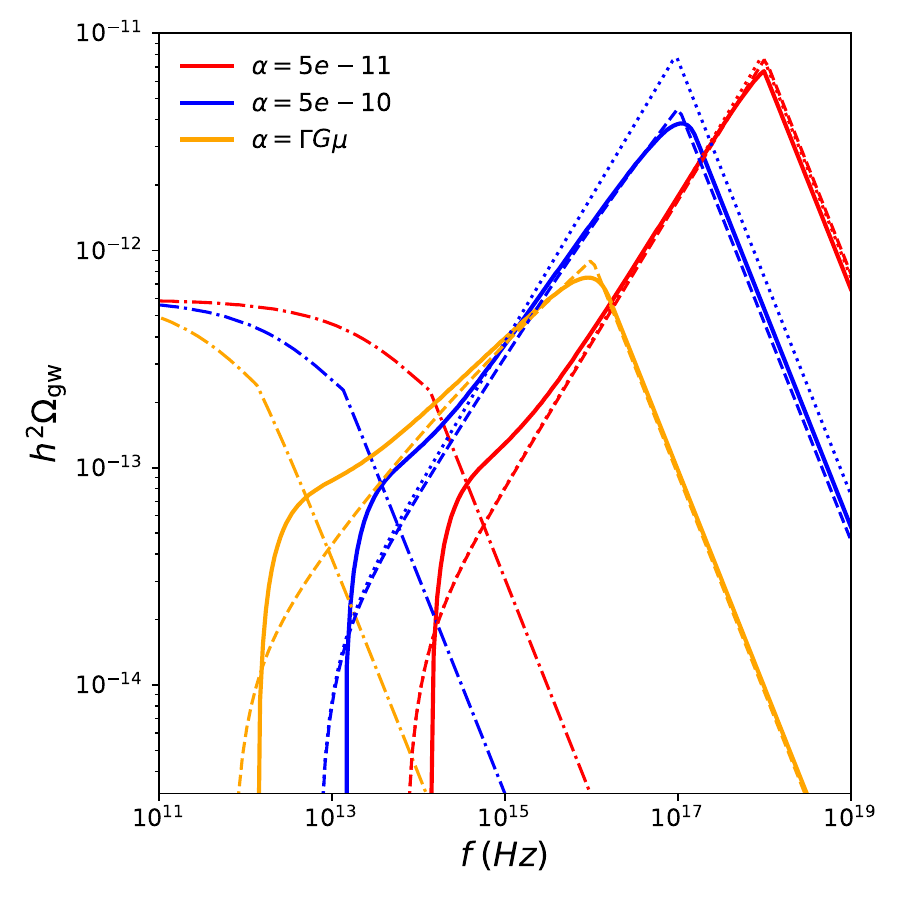}
     \caption{Analytical approximation to the SGWB generated by cosmic string loops created during the friction dominated epoch in the transitional regime. Here, we took $G \mu=10^{-10}$, $\beta=10$, $L_c=t_c$ and considered different values of $\alpha$ satisfying $0.01\,\Gamma G\mu \lesssim\alpha\lesssim\Gamma G\mu$. The solid lines represent spectra computed numerically, while the dotted and dashed lines correspond to the analytical approximations in Eqs.~(\ref{SGWB_an_sm_loops}) and~(\ref{SGWB_an_back_more}) respectively. The dash-dotted lines represent the contribution from the frictionless period of the networks evolution.}
     \label{fig:SGWB_analytically_appendix}
\end{figure}

We have also thoroughly tested the validity of this approximation for the range of parameters in Eq.~\eqref{ranges_of_parameters}, but restricted $\alpha$ to $0.01\,\Gamma G\mu \lesssim\alpha\lesssim\Gamma G\mu$ (i.e. the range where other approximations lose quality). We have verified that it provides an excellent fit in this transitional range. We display an illustrative example of this approximation in Fig.~\ref{fig:SGWB_analytically_appendix}, where the dashed and dotted lines correspond to $\Omega^{\rm tr}_{\rm gw}$ and $\Omega^{\rm sm}_{\rm gw}$, respectively. Therein, one may see that it describes very well the signature of friction for $\alpha = 0.1\, \Gamma G\mu$ and $\alpha = 0.01\, \Gamma G\mu$ (the blue and red lines, respectively), which generally cannot be described by $\Omega^{\rm sm}_{\rm gw}$ and $\Omega^{\rm br}_{\rm gw}$. For $\alpha = \Gamma G\mu$, $\Omega^{\rm tr}_{\rm gw}$ differs slightly from $\Omega^{\rm br}_{\rm gw}$ (cf. Fig.~\ref{fig:SGWB_analytically_large_1}) but nevertheless provides a good description of the signature too. From Eqs.~(\ref{SGWB_an_back_simple}) and~(\ref{SGWB_an_back_more}), it is straightforward to see that the solution $\Omega^{\rm tr}_{\rm gw}$ encompasses $\Omega^{\rm br}_{\rm gw}$. As a result, for $\alpha > \Gamma G\mu$ the two approximations coincide. Similarly, for $\alpha < 0.01\, \Gamma G\mu$, $\Omega^{\rm tr}_{\rm gw}$ matches $\Omega^{\rm sm}_{\rm gw}$ in the vast majority of cases.

\bibliography{approx}
\end{document}